\documentclass[12pt]{article}

\usepackage{latexsym} 
\usepackage{amsmath}
\usepackage{amssymb}
\usepackage{graphicx}
\usepackage{wrapfig}
\usepackage{sectsty}
\usepackage{indentfirst}

\allsectionsfont{\centering}
\makeatletter

\newcommand{\ctg}{\cot}
\newcommand{\tot}{\mathop{\rm tot}\nolimits}

\def\s#1{\sqrt{#1}}

\newcommand{\be}{\begin{equation}}
\newcommand{\ee}{\end{equation}}
\newcommand{\ba}{\begin{eqnarray}}
\newcommand{\ea}{\end{eqnarray}}
\newcommand{\pa}{\partial}
\let\f\frac

\newcommand{\ds}{\displaystyle}

\begin{document}

\begin{center}
\large
{\Large\textbf{{\textit{Self-restriction} of Gravitational
Field
and its Role in the Universe}}}

\bigskip

S.S. Gershtein,  A.A.
Logunov\footnote{e-mail: Anatoly.Logunov@ihep.ru} \\and M.A.
Mestvirishvili
             
\vspace*{0.2cm}
{\small{\it Institute for High Energy Physics, Protvino, Russia}}

\end{center}

\vspace*{0.5cm}
\centerline{Abstract}
\vspace*{0.2cm}
{\small
It is shown in the article that according to the Relativistic Theory
of Gravitation the gravitational field providing slowing down of the
time 
rate nevertheless stops  itself this slowing down in  strong fields.
So a physical tendency of this field to
\textit{self-restriction} of the gravitational potential is 
demonstrated. 
This property of the field leads to a stopping of the collapse of
massive bodies and to the cyclic
evolution of the
homogeneous and isotropic Universe.}

\section*{Introduction}

Both in the Newton gravity theory and in the Einstein general theory
of relativity (GTR) the gravitational forces are attractive forces.
However the field approach to gravitation shows that it is not quite
so in strong gravitational fields. We shall return to this point
later. 

The Relativistic Theory of Gravitation (RTG) was presented in detail
on the pages of ``Uspekhi'' in paper~[1]. Here we only briefly enlist
its fundamental pro\-po\-si\-ti\-ons.

In the basis of the RTG lies the special theory of relativity and
this provides energy-momentum and angular momentum conservation
laws for all physical pro\-ces\-ses including the gravitational ones.
The 
RTG stems from a hypothesis that gravitation is universal and its
source is the conserved energy-momentum tensor of all matter fields
including that of the gravitational field. 

That is why the gravitational field is a tensor field, 
$\phi^{\mu\nu}$.  
Such an approach corresponds to Einstein's idea that he wrote about
as early as in 1913~[2]:
\textit{``\ldots the tensor of the gravitational field
$\vartheta_{\mu\nu}$
is a source of the field equality with the material systems tensor 
$\Theta_{\mu\nu}$. Exeptional status of the energy of the
gravitational field in comparison with other kinds of energy would
leads to inadmissible consequences
''.}
This Einstein's idea was put into the basis of the relativistic
theory of gravitation. Einstein, when devising the general theory of
relativity, did not manage to realize this idea because instead of
the
energy-momentum tensor a pseudo-tensor of the gravitational field
arose in the GTR. This happened because Einstain did not consider the
gravitational field as a physical field (of a Faradey--Maxwell type)
in the Minkowski space. That is why the GTR equations do not contain
the metric of the Minkowski space. The approach to gravitation
adopted in the RTG implies 
\textit{geometrization}, i.e. an effective Riemannian space results 
\textit{but of a simple topology only}. This leads to the folowing
picture: the test body motion in the Minkowski space under the action
of the gravitational field is equivalent to the motion of this body
in an effective Riemannian space created by the gravitational field.
Namely this circumstance of the theory allows one to separate
inertial forces from the gravitational ones. 

In the field approach to gravitation an effective Riemannian space
arises but of a simple topology only. This is the reason why the
field approach cannot lead to the GTR in which the topology is not
simple in general. 

This idea described above leads to the following complete system of
equations 
[1, 3, 4]:
\be
\Bigl(R^{\mu\nu}-\f{\,1\,}{2}g^{\mu\nu}R\Bigr)+\f{m^2}{2}
\Bigl[g^{\mu\nu}+\Bigl(g^{\mu\alpha}g^{\nu\beta}
-\f{\,1\,}{2}g^{\mu\nu}g^{\alpha\beta}\Bigr)\gamma_{\alpha\beta}\Bigr
]
=8\pi GT^{\mu\nu}\,,
\label{eq1'}
\ee
\be
D_\nu\tilde{g}^{\nu\mu}=0\,.
\label{eq2'}
\ee
Here $T^{\mu\nu}$  is the substance energy-momentum tensor
\footnote{Under the ``substance'' we mean all physical fields except
the gravitational one.}, 
$D_\nu$ stands for the covariant derivative in the Minkowski space,
$\gamma_{\alpha\beta}$ is the metric of the Minkowski space;
$g_{\alpha\beta}$ is the metric of the effective Riemannian space,  
$m=m_gc/\hbar$, $m_g$ is the graviton mass, 
$\tilde{g}^{\nu\mu}=\sqrt{-g}\,g^{\nu\mu}$  is the density  of the
metric tensor  
$g^{\mu\nu}$.

Due to the non-zero mass at rest of the graviton in equation
(\ref{eq1})
equations 
(\ref{eq2}) are a consequence of the gravitational field equations 
(\ref{eq1}) and the substance equaitons.

The effective metric of the Riemannian space 
$g^{\mu\nu}$ is related to the gravitational field 
$\phi^{\mu\nu}$ by the relationship
\[
\tilde{g}^{\mu\nu}=\tilde{\gamma}^{\mu\nu}+\tilde{\phi}^{\mu\nu}\,,
\]
where
\[
\tilde{\gamma}^{\mu\nu}=\sqrt{-\gamma}\,\gamma^{\mu\nu},\quad
\tilde{\phi}^{\mu\nu}=\sqrt{-\gamma}\,\phi^{\mu\nu},\quad
\gamma =\mbox{det}\,\gamma_{\mu\nu}\,.
\]

The system of equations (\ref{eq1'}), (\ref{eq2'}) is generally
covariant
under arbitrary co\-o\-r\-di\-na\-te transformations and
form-invariant under
the Lorentz transformations. It follows directly from the least
action principle with a Langrangean density 
\[
L=L_g(\gamma_{\mu\nu},\tilde{g}^{\mu\nu})
+L_M(\tilde{g}^{\mu\nu},\phi_A)\,,
\]
here
\[
L_g=\f{1}{16\pi}\tilde{g}^{\mu\nu}(G_{\mu\nu}^\lambda
G^\sigma_{\lambda\sigma} -G_{\mu\sigma}^\lambda
G^\sigma_{\nu\lambda})
-\f{m^2}{16\pi}\Bigl(\f{\,1\,}{2}\gamma_{\mu\nu}\tilde{g}^{\mu\nu}
-\sqrt{-g}-\sqrt{-\gamma}\Bigr)\,,
\]
\[
G_{\mu\nu}^\lambda =\f{\,1\,}{2}g^{\lambda\sigma}
(D_\mu g_{\sigma\nu}+D_\nu g_{\sigma\mu}-D_\sigma g_{\mu\nu})\,,
\quad \phi_A\, {\mbox{ are\, the\, substance\,fields}}\,.
\]

In order that timelike and isotropic intervals in the effective
Riemannian space could not get out of the light-cone of the
underlying
Minkowski space the causality condition has to hold 
\be
\gamma_{\mu\nu}v^\mu v^\nu =0\,,\quad
g_{\mu\nu}v^\mu v^\nu\le 0\,,
\label{eq3}
\ee
Here $v^\nu$ is the velocity four-vector.

Thus the motion of test bodies under the action of the gravitational
field proceeds always 
\textit{inside} both the Riemannian cone and the cone of the
Minkowski space. This provides the geodesic completness. 

The rest mass of the graviton arises unavoidably in the theory
because only in this way one can consider the gravitational field as
a physical field in the Minkowski space whose sourse is the total
conserved energy-momentum tensor of all matter. And this is a
non-zero
mass of the graviton that changes completely the picture both of the
collapse process and the evolution of the Universe. 

When A.~Einstein in 1913 related the gravitational field to the
metric tensor of the Riemannian space it appeared that such a field
caused a slowing down of the lapse of a physical process. 
This slow down can be illustrated, in particular, in the case of the
Schwarzschild solution, if to compare the lapse of time in the
presence
of the gravitational field with the lapse of time for a distant
observer. However, generally, only the metric tensor of the
Riemannian space takes place in the GTR and therefore any trace of
inertial time of the Minkowski space is absent from the
Hilbert--Einstein equations. Due to this reason the universal
property of the gravitational field to slow down the lapse of time in
comparison with the inertial time could not get a further development
in the framework of the GTR.

The rise of the effective Riemannian space in the field theory of
gravitation with preservation of the Minkowski space as a basic space
gives  a special importance to the property of the gravitational
field to exert a slowing down influence on the lapse of time. 
Only in this case can one  argue truly about the 
 \textit{slowing down of the lapse of time when making the comparison
 of the lapse of time in the gravitation field with that in an
 inertial
 frame of the Minkowski space in the absence of gravitation}.

All this is realized in the RTG because the metric tensor
$\gamma_{\mu\nu}$ of the Minkowski space enters  explicitly the full
system of its equations.

To demonstrate that the change of the lapse of time implies the
appearance of a force we turn to the Newton equation
\[
m\f{d^2x}{dt^2}=F\,.
\]
If one passes formally from the inertial time to a time 
$\tau$ with 
\[
d\tau =U(t)dt,
\]
then it is easy to obtain 
\[
m\f{d^2x}{d\tau^2}=\f{1}{U^2}\Bigl\{F-\f{dx}{dt}\f{d}{dt}\ln
U\Bigr\}\,.
\]
One can see from this that the change of the lapse of time defined by
the function  $U$ results in the appearance of an effective force. 
All this bears a formal  character here as in this case there is
no  physical reason which would change the lapse of time. But this
formal example shows that if a process of slowing down of the lapse
of time occurs is \textit{Nature} then it unavoidably generates
effective field forces, and so it is necessary to take them into
account as something absolutely 
new and surprising. The physical gravitational force changes both the
lapse of time and parameters of the space quantities in comparison
with
the same quantities is an inertial system of the Minkowski space
without gravitation. 

The field approach to gravitation excludes the concept of black holes
and explaines the evolution both of massive bodies and the Universe
on the basis of more profound insight into the physical properties of
the very gravitational field.  

This confirms the deep intuition of  A.\,S.\, Eddington who said at
the session of the Royal Astronomical Society  11 January 1935:
\textit{``The star has to go on radiating and radiating and
contracting and contracting until, I suppose, it gets down to a few
km. radius, when gravity becomes strong enough to hold in the
radiation, and the star can at last find peace. \ldots I felt driven
to the conclusion that this was almost \textbf{a reductio ad
absurdum}
of the relativistic degeneracy formula. Various accidents may
intervene to save the star, but I want more protection than that. I
think
there should be a law of Nature to prevent a star from behaving in
this absurd way\ !''}~\footnote{The
Observatory. 1935.
Vol.~58. P.~373.}

It appears that in the framework of the field formulation of
gravitation such a 
\textit{Law of Nature} is contained in the physical property of the
gravitational field to stop the 
 \textit{process} of the slowing down of the lapse of time and hence
 to
 limit its potential. This stops the 
\textit{process}  of compression. 

Below, taking as examples the collapse and the evolution of the
homogeneous and isotropic Universe,  we will see in what way the
\textit{self-restriction} of the gravitational field potential arises
which stops both the process of the slowing down of time and the
process of the substance compression.

\section{Equations of the spherically symmetric static gravitational~field}

The interval in the Minkowski space has the following form in
spherical coordinates
\be
d\sigma^2 =(dx^0)^2
-(dr)^2-r^2 
(d\theta^2+\sin^2\theta\,d\phi^2)\,,
\label{eq4}
\ee
here $x^0=ct$. The interval in the effective Riemannian space for
a spherically symmetric static field can be written in the form
\be
ds^2 =U(r)(dx^0)^2-V(r)dr^2-W^2(r) 
(d\theta^2+\sin^2\theta\,d\phi^2)\,.
\label{eq5'}
\ee
Equations  (\ref{eq1'}), ({\ref{eq2'}) of the RTG we represent in the
form
\be
R^\mu_\nu -\f{\,1\,}{2}\delta^\mu_\nu R+\f{m^2}{2}
\Bigl(\delta^\mu_\nu +g^{\mu\alpha}\gamma_{\alpha\nu}
-\f{\,1\,}{2}\delta^\mu_\nu g^{\alpha\beta}\gamma_{\alpha\beta}\Bigr)
=\varkappa T^\mu_\nu\,,  
\label{eq6'}
\ee
\be
D_\mu\tilde g^{\mu\nu}=0\,. 
\label{eq7'}
\ee
Equation  (\ref{eq7'}) reads in more detail
\be
\pa_\mu\tilde g^{\mu\nu}+\gamma_{\lambda\sigma}^\nu\tilde
g^{\lambda\sigma}=0\,. \label{eq8'}
\ee
Here $\gamma_{\lambda\sigma}^\nu$ are Christoffel's symbols of the
Minkowski space. For a spherically symmetric static source the
components of the tensor 
$T^\mu_\nu$ are:
\be
T_0^0=\rho(r),\quad T_1^1=T_2^2=T_3^3=-\f{p(r)}{c^2}\,,
\label{eq9'}
\ee
here $\rho$ is the mass density, $p$ stands for the isotopic
pressure. 

For the definition of the metric coefficients  $U, V$ and $W$ one can
make use of Eqs.(\ref{eq6'}) for 
$\mu =0,\, \nu=0;\, \mu=1, \nu=1$.
\ba
&\ds\f{1}{W^2}-\f{1}{VW^2}
\Bigl(
\ds\f{dW}{dr}\Bigr)^2
-\ds\f{2}{VW}\f{d^2W}{dr^2}
-\ds\f{1}{W}\f{dW}{dr}\f{d}{dr}
\Bigl(\f{1}{V}\Bigr)+\nonumber \\[2mm]
&+\ds\f{\,1\,}{2}m^2\Bigl[1+\ds\f{\,1\,}{2}
\Bigl(\ds\f{1}{U}-\ds\f{1}{V}\Bigr)
-\ds\f{r^2}{W^2}\Bigr]
=\varkappa \rho\,,\label{eq10'} 
\ea
\ba
&\ds\f{1}{W^2}-\f{1}{VW^2}
\Bigl(\f{dW}{dr}\Bigr)^2
-\f{1}{UVW}\f{dW}{dr}\f{dU}{dr}+\nonumber \\[2mm]
&+\ds\f{\,1\,}{2}m^2\Bigl[1-\f{\,1\,}{2}
\Bigl(\f{1}{U}-\f{1}{V}\Bigr)
-\f{r^2}{W^2}\Bigr]=-\varkappa \f{p}{c^2}\,.\label{eq11'} 
\ea

Equation (\ref{eq8'}) takes the form
\be
\f{d}{dr}\Bigl(\sqrt{U/V}\,W^2\Bigr)
=2r\sqrt{UV}\,.
\label{eq12'}
\ee
With account of the identity
\[
\f{dr}{dW}\f{1}{W^2}\f{d}{dr}
\Bigl[\f{W}{V}\Bigl(\f{dW}{dr}\Bigr)^2\Bigr]=
\f{1}{VW^2}\Bigl(\f{dW}{dr}\Bigr)^2
+\f{2}{VW}\f{d^2W}{dr^2}
+\f{1}{W}\f{dW}{dr}\f{d}{dr}\Bigl(\f{1}{V}\Bigr)\,,
\]
and after passing from the derivatives in  $r$ to the derivatives in
$W$ equations (\ref{eq10'}), (\ref{eq11'}) and (\ref{eq12'}) take the
form
\be
1-\f{d}{dW}\Bigl[\f{W}{V(dr/dW)^2}\Bigr]
+\f{\,1\,}{2}m^2\Bigl[W^2-r^2+\f{W^2}{2}
\Bigl(\f{1}{U}-\f{1}{V}\Bigr)\Bigr]=\varkappa W^2\rho\,,
\label{eq13'}
\ee
\ba
&1-\ds\f{W}{V(dr/dW)^2}\f{d}{dW}[\ln (UW)]+\nonumber \\[2mm]
&+\ds\f{\,1\,}{2}m^2\Bigl[W^2-r^2-\f{W^2}{2}
\Bigl(\f{1}{U}-\f{1}{V}\Bigr)\Bigr]=-\varkappa 
W^2\f{p}{c^2}\,,\label{eq14'} 
\ea
\vspace*{2mm}
\be
\f{d}{dW}\bigl[\sqrt{U/V}\,W^2\bigr]=2r\sqrt{UV}\,\f{dr}{dW}\,.
\label{eq15'}
\ee
Further on we will use these equations for various state equations of
the substance. Namely on the basis of these equations it will be
demonstrated in Parts~2,3 and 4 that the gravitational field
possesses a property of  
 \textit{ self-restriction} which imposes the limit for the slowing
 down
 of the lapse of time by the gravitational field. 

\section{ External solution for the spherically symmetric
static body}

In this Part it will be shown that presence of the rest mass of the
graviton changes qualitatively the character of the solution in the
region near to the Schwar\-zschild sphere. Below we consider this
in
detail. 

Substracting Eq.(\ref{eq14'}) from (\ref{eq13'}) and introducing new
variables
\be
Z=\f{UW^2}{V\dot{r}^2},\quad
\dot{r}=\f{dr}{dt},\quad
t=\f{W-W_0}{W_0}\,,
\label{eq16}
\ee
we obtain
\be
\f{dZ}{dW}-\f{2Z}{U}\f{dU}{dW}-2\f{Z}{W}
-\f{m^2W^3}{2W_0^2}
\Bigl(1-\f{U}{V}\Bigr)
=-\varkappa \f{W^3}{W_0^2}\Bigl(\rho +\f{p}{c^2}\Bigr)U\,.
\label{eq17}
\ee
Adding Eqs.(\ref{eq13'}) and (\ref{eq14'}), we find 
\be
1-\f{\,1\,}{2}\f{W_0^2}{W}\f{1}{U}\f{dZ}{dW}
+\f{m^2}{2}(W^2-r^2)
=\f{\,1\,}{2}\varkappa W^2\Bigl(\rho -\f{p}{c^2}\Bigr)\,.
\label{eq18}
\ee
Let us consider (\ref{eq17}) and (\ref{eq18}) outside the substance
in the region defined by the inequalities
\be
\f{U}{V}\ll 1,\quad \f{\,1\,}{2}m^2(W^2-r^2)\ll 1\,.
\label{eq19}
\ee
In this region Eq.(\ref{eq18}) has the form
\be
U=\f{\,1\,}{2}\f{W_0^2}{W}\f{dZ}{dW}
=\f{\,1\,}{2}\f{W_0}{W}\f{dZ}{dt}\,.
\label{eq20}
\ee
Taking into account  (\ref{eq20}) we bring Eq.(\ref{eq17}) to the
form 
\be
Z\f{d^2Z}{dW^2}-\f{\,1\,}{2}\Bigl(\f{dZ}{dW}\Bigr)^2
+\f{\,1\,}{4}m^2\f{W^3}{W_0^2}\f{dZ}{dW}=0\,.
\label{eq21}
\ee
Let us introduce, according (\ref{eq16}), variable $t$. Then
Eq.(\ref{eq21}) assumes the  form
\be
Z\ddot{Z}-\f{\,1\,}{2}(\dot{Z})^2
+\alpha (1+t)^3\dot{Z}=0\,,
\label{eq22}
\ee
Here $\alpha =m^2W_0^2/4,\;\dot{Z}=dZ/dt$.
For  $t$ defined by the inequality
\be
0\leq t\ll 1/3\,,
\label{eq23}
\ee
Eq.(\ref{eq22}) simplifies
\be
Z\ddot{Z}-\f{\,1\,}{2}(\dot{Z})^2
+\alpha\dot{Z}=0\,.
\label{eq24}
\ee
It has a solution 
\be
\lambda\sqrt{Z}=2\alpha\ln\Bigl(1+\f{\lambda\sqrt{Z}}{2\alpha}\Bigr)
+\f{\lambda^2}{2}t\,. 
\label{eq25}
\ee
Here $\lambda$ is an arbitrary constant. 

On the basis of  (\ref{eq20}) and (\ref{eq16}) we have 
\be
U=\f{\,1\,}{2}\f{W_0}{W}\dot{Z},\quad
V\dot{r}^2=\f{\,1\,}{2}W_0W\f{\dot Z}{Z}\,.
\label{eq26}
\ee
Making use of  (\ref{eq25}) we find
\be
\dot{Z}=2\alpha +\lambda\sqrt{Z}\,.
\label{eq27}
\ee
Substituting  (\ref{eq27}) into (\ref{eq26}), we obtain
\be
U=\f{W_0}{W}\Bigl(\alpha +\f{\lambda}{2}\sqrt{Z}\Bigr),\quad
V\dot{r}^2=W_0W\,\f{\alpha +\lambda\sqrt{Z}/2}{Z}\,.
\label{eq28}
\ee
At $\alpha =0$ we have from  (\ref{eq25}) 
\be
\sqrt{Z}=\f{\,\lambda\,}{2}t\,.
\label{eq29}
\ee
Substituting this expression into (\ref{eq28}),  we find
\be
U=\Bigl(\f{\,\lambda\,}{2}\Bigr)^{\!2}\,\f{W-W_0}{W}\,.
\label{eq30}
\ee
But this expression for $U$ has to coincide exactly with the
Schwarzschlid solution 
\be
U=\f{W-W_g}{W},\quad W_g=\f{2GM}{c^2}\,.
\label{eq31}
\ee
Comparing  (\ref{eq30}) and (\ref{eq31}), we obtain
\be
\lambda =2,\quad W_0=W_g\,.
\label{eq32}
\ee
Thus we find:
\be
U=\f{W_g}{W}(\alpha +\sqrt{Z}),\quad V\dot{r}^2=W_g W\f{\alpha
+\sqrt{Z}}{Z}\,.
\label{eq33}
\ee
We need now to define how  $r$ depends on $W$ with  help of (15)
Substituting  (\ref{eq33}) into Eqs.(15) and passing to the
variable
\be
\ell =r/W_g\,,
\label{eq34}
\ee
we obtain
\be
\f{d}{d\sqrt{Z}}\Bigl[(1+t)\f{dZ}{dt}\f{d\ell}{d\sqrt{Z}}\Bigr]=4\ell
\,.
\label{eq35}
\ee
Taking into account  (\ref{eq27}) and making differentiation in
$\sqrt{Z}$ in (\ref{eq35}),  we find 
\be
(1+t)(\alpha+\sqrt{Z})\f{d^2\ell}{(d\sqrt{Z})^2}+
(1+t+\sqrt{Z})\f{d\ell}{d\sqrt{Z}}-2\ell =0\,.
\label{eq36}
\ee
As we are interested in the region of  $t$ defined by inequality 
(\ref{eq23}), Eq.(\ref{eq36}) is simplified and has the form
\be
(\alpha+\sqrt{Z})\f{d^2\ell}{(d\sqrt{Z})^2}+
(1+\sqrt{Z})\f{d\ell}{d\sqrt{Z}}-2\ell =0\,.
\label{eq37}
\ee
The general solution of Eqs.(\ref{eq37})  will be
\be
\ell=A\ell_1+B\ell_2\,,
\label{eq38}
\ee
where
\[
\ell_1 =F[-2,\,1-\alpha,\,-(\alpha +\sqrt{Z})],\;
\ell_2 =(\alpha+\sqrt{Z})^\alpha F[-2+\alpha,\,1+\alpha,\,-(\alpha
+\sqrt{Z})]\,.
\]
Here  $A$ and $B$  are arbitrary constants, 
$F$ is a degenerated hypergeometric function. 

The analysis of solution  (\ref{eq38}) in the region defined by
inequalities  (\ref{eq19}) and (\ref{eq23})  leads to an equality
\be
\dot{r}=W_g\,.
\label{eq39}
\ee
Let us consider the following limiting case
\be
\sqrt{Z}\gg\alpha\,.
\label{eq40}
\ee
In this case we have from expression  (\ref{eq25}) with account of
(\ref{eq32}) 
\be
\sqrt{Z}=t\,.
\label{eq41}
\ee
Substituting this expression into  (\ref{eq28}) and taking into
consideration   (\ref{eq32}), (\ref{eq39}),  we obtain the
Schwarzschild solution
\be
U=\f{W-W_g}{W},\quad V=\f{W}{W-W_g}\,.
\label{eq42}
\ee

Now we come to another limiting case when the influence of the
graviton mass is essential, 

Let the following inequality takes place 
\be
\sqrt{Z}\ll\alpha\,.
\label{eq43}
\ee
In this approximation we find from expression (\ref{eq25})  with
account of (\ref{eq32})
\be
Z=2\alpha t\,.
\label{eq44}
\ee
Substituting this expression into  (\ref{eq28}) and taking into
consideration  (\ref{eq32})  and 
(\ref{eq39}),  we obtain
\be
U=\alpha\f{W_g}{W},\quad V=\f{\,1\,}{2}\f{W}{W-W_g}\,.
\label{eq45}
\ee
This solution, on the basis of (\ref{eq43})   and (\ref{eq44}),  
holds in the region
\[
t\ll\f{\alpha}{2},\;\;\mbox{i.\,e.}\;\;
W-W_g\ll\f{\,1\,}{2}W_g\Bigl(\f{m_gc}{\hbar}\,\f{W_g}{2}\Bigr)^2 \ .
\]
We see from  (\ref{eq45}), that the graviton mass 
$m_g$  does not allow for vanishing of the quantity 
$U$. \textit{The rest mass of the graviton imposes for any body its
limit for slowing-down of the lapse of time}. This limit is defined
by
a linear function of the Schwarzschild radius, i.e. of the mass of
the body
\[
\f{\,1\,}{2}\Bigl(\f{m_gc}{\hbar}\Bigr)W_g\,.
\]
There is no such a limit in the GTR. Such a property of the
gravitational field leads to a cardinal change in the motion of a
test body in the gravitational field. 

The motion of a test body proceeds along a geodesic line of the
Riemannian space
\be
\label{eq46}
\f{dv^\mu}{ds}+\varGamma_{\alpha\beta}^\mu
\f{dx^\alpha}{ds}\cdot\f{dx^\beta}{ds}=0\,.
\ee 
Here $v^\mu=dx^\mu /ds$ is the velocity four-vector, and  $v^\mu$
satisfies the condition
\be
\label{eq47}
g_{\mu\nu}v^\mu v^\nu =1\,.
\ee 

Let us consider the radial motion 
\be
\label{eq48}
v^\theta =v^\phi =0,\quad v^r =dr/ds\,.
\ee 
Taking into consideration that the Christoffel symbol
$\varGamma_{01}^0$ is 
\be
\label{eq49}
\varGamma_{01}^0=\f{1}{2U}\f{dU}{dr}\,,
\ee 
we find from Eq. (\ref{eq46}) 
\be
\label{eq50}
\f{dv^0}{ds}+\f{1}{U}\f{dU}{dr}v^0 v^r =0\,.
\ee 
Resolving equation  (\ref{eq50}), we obtain 
\be
\label{eq51}
\f{d}{dr}\ln (v^0 U) =0\,.
\ee 
We have thereof:
\be
\label{eq52}
v_0=\f{dx^0}{ds}=\f{U_0}{U}\,,
\ee 
$U_0$  being an integration constant. If to assume the velocity of
the falling body equal to zero at infinity we obtain 
$U_0=1$. From relation (\ref{eq47}) we find 
\be
\label{eq53}
\f{dr}{ds}=-\sqrt{\f{1-U}{UV}}\,.
\ee 
Substituting this into expression  (\ref{eq45}) and taking into
consideration  (\ref{eq39}), we obtain 
\be
\label{eq54}
\f{dW}{ds}=-\Bigl(\f{\hbar}{m_g c}\Bigr)\f{2}{W_g}
\sqrt{2\f{W}{W_g}\Bigl(1-\f{W_g}{W}\Bigr)}\,.
\ee 
It is seen thereof that a turning point appears. Differentiating  
(\ref{eq54}) with respect to  $s$, we find  
\be
\label{eq55}
\f{d^2 W}{ds^2}=4\Bigl(\f{\hbar}{m_g c}\Bigr)^2\f{1}{W_g^3}\,.
\ee 
We see that in the turning point the acceleration is positive, i.e. a
repulsion takes place, and  it is significant. Integrating 
(\ref{eq54}) we obtain
\be
\label{eq56}
W=W_g+2\Bigl(\f{\hbar}{m_g c}\Bigr)^2\f{(s-s_0)^2}{W_g^3}\,.
\ee 
It is clear from expression  (\ref{eq56}) that  
\textit{a test body cannot cross} the Schwarzschild sphere. 

According to expressions (\ref{eq45}) the scalar quantity 
$g/\gamma$, where $g=\mbox{det}$ $g_{\mu\nu}$,
$\gamma =\mbox{det}$ $\gamma_{\mu\nu}$, 
has a singularity at 
$W=W_g$ that cannot be eliminated by the choice of the coordinate
system. Therefore the presence of such a singularity in vacuum is
inadmissible, otherwise one cannot sew up external solution with the
solution inside the body. The conclusion follows from this that the
body radius has to be larger than the Schwarzschild radius. In such a
way a 
\textit{self-restriction} of the field strength arises in the RTG and
so the very reason of the appearance of the ``Schwarzschild
singularity'' dissapears. This corresponds completely to A.~Einstein
opinion that he expressed as early  as in 1939 in paper~[5]:
\textit{``The main result of the conducted study is a \textbf{clear
understanding that ``Schwarzschild singularities'' are absent from
the real world} (emphasized by us. Authors)''} 
And further on:
\textit{``The Schwarzschild singularity is absent because one cannot
concentrate the substance in an arbitrary way, otherwise the
particles forming the clusters will \textbf{achieve the velocity of
light} (emphasized by us. Authors)''}.

As an example we consider gravitational field in a shrinking
(synchronous) coordinate system. The passage to this coordinate
system from an inertial one is being made by means of transformations
\[
dt=\f{\,1\,}{U}[d\tau -dR(1-U)],\quad 
dW=\sqrt{\ds\f{1-U}{UV}}(dR-d\tau)\,.
\]
In the synchronous coordinate system the interval of the Riemannian
and pseudo-Euclidean space-times has the form
\[
ds^2 =d\tau^2
-[1-U(X)]dR^2
-W^2(X)(d\theta^2+\sin^2\theta\,d\phi^2)\,,
\]
\ba
&&d\sigma^2=d\tau^2\,\ds\f{1-\dot{r}^2U^2}{U^2}
+2\,dR\,d\tau\,\ds\f{\dot{r}^2U^2-(1-U)}{U^2}-\nonumber\\
&&\qquad\!\! -\,dR^2\,\ds\f{\dot{r}^2U^2-(1-U)^2}{U^2}
-r^2(d\theta^2+\sin^2\theta\,d\phi^2)\,,\nonumber
\ea
where  $X=R-\tau,\;\dot{r}=dr/dX$.

The  RTG equations 
$$
R_{\mu\nu}=8\pi G\Bigl(T_{\mu\nu}-\f{\,1\,}{2}g_{\mu\nu}T\Bigr)
+\f{m^2}{2}(g_{\mu\nu}-\gamma_{\mu\nu})\,,
\eqno{(a)}
$$
$$
D_\nu\tilde{g}^{\mu\nu}=0\,,
$$
for the problem defined by the intervals  $ds^2$ and $d\sigma^2$ lead 
(outside the substance) to the equations of the form
$$
R_{01}=\f{2\ddot{W}}{W}
+\f{1}{(1-U)W}\dot{U}\dot{W}
=\f{m^2}{2}\Bigl(\f{1-U}{U^2}-\dot{r}^2\Bigr)\,,
\eqno{(b)}
$$ 
$$
R_{00}+R_{01}=\f{1}{1-U}
\Bigl[\f{\,1\,}{2}\ddot{U}
+\f{\dot{U}^2}{4(1-U)}
+\f{1}{W}\dot{U}\dot{W}\Bigr]
=-\f{m^2}{2}\f{1-U}{U}\,,
\eqno{(c)}
$$
In the region of the variable $X$ where one can neglect the graviton
mass due to its smallness we find from these equations
$$
W=W_g^{1/3}
\Bigl[\f{\,3\,}{2}X\Bigr]^{2/3}\!,\quad
1-U=\Bigl[\f{\,2\,}{3}W_g\Bigr]^{2/3}X^{-2/3}\,,
\eqno{(d)}
$$
From expression  (d) for the  function  $U$ it follows that it
decreases
with decreasing of 
$X$, and its derivative  $\dot{U}$  is positive. 

In approximation   (\ref{eq19}) we find from equation $(a)$
outside the substance 
\[
R_{22}=-\f{UW}{1-U}\ddot{W}
-\f{U}{1-U}\dot{W}^2
-\f{W(2-U)}{2(1-U)^2}\dot{U}\dot{W}+1=0\,.
\]
In the region of small values  $0<U\ll 1$ the equation is somewhat
simplified and assumes the form
\[
UW\ddot{W}
+U\dot{W}^2
+W\dot{U}\dot{W}-1=0\,.
\]
This equation has a solution
\[
\dot{W}=\f{X}{UW}\,.
\]
At the stopping point
\[
\dot{W}=0\,,
\]
the second derivative $\ddot{W}$ at small values of 
$U$ is positive according to equations 
$(b)$ and $(c)$ and this is an evidence of the presence of a
repulsive
force. 

It  is this point where the process of expansion starts from. This
expansion stops is the region of 
$X$where equalities  $(d)$ hold. In this region 
$\ddot{W}$ is negative 
\[
\ddot{W}=-\f{\,1\,}{2}W_g^{1/3}\Bigl[\f{\,3\,}{2}X\Bigr]^{-4/3},
\]
and consequently attraction takes place. So if the stopping point
were
outside the substance then  compression would start after expansion,
then again stop and again expansion and so on. 
However the real
gravitational field excludes such a regime. 

While in the GTR for the given problem an equation takes place 
\[
W=\Bigl[\f{\,3\,}{2}\bigl(R-c\tau\bigr)\Bigr]^{2/3}W_g^{1/3},
\]
in our case we obtain the expression 
\[
W=W_g+2\Bigl(\f{\hbar}{m_g c}\Bigr)^2\f{(R-c\tau)^2}{W_g^3}\,,
\] 
that excludes achieving the point  $W=0$.
This means that 
\[
\f{d^2W}{d\tau^2}=\f{4c^2}{W_g^3}\Bigl(\f{\hbar}{m_g c}\Bigr)^2. 
\]
All this occurs because of the passage from the inertial time 
$t$ to the physical time  $\tau$.

As the gravitational field is created by the substance and the very
gravitational field restricts its potential, it follows from the
example above that for obtaining a physical solution one needs to sew
up  the solution inside the substance with the external solution but
it is necessary that the gravitational field potential would be
bounded on the surface of the body by the inequality
\[
\f{|\phi|}{c^2}<1\,.
\]

It is such a solution, which corresponds to the real gravitational
field, 
that implies that the stoppping point cannot lie in vacuum. That is
why
the world lines of particles which are at rest in the shrinking
coordinate
system will collide with the substance of the source of the field,
and
besides these collisions will occur during a finite time for any
observer. All this excludes the regime of motion about  which we
wrote  
above. At the same time this  excludes the appearance of ``black
holes''. 

Let us turn now to analysis of the internal solution.
\vspace*{-2mm}

\section{ Internal solution of the Schwarzschild type}
\vspace*{-2mm}

In paper [6]  Schwarzschild has found a spherically symmetric static
internal solution of the equations of the general theory of
relativity. For a  \textit{homogeneous ball}
of radius  $a$ it is described by the interval 
\ba
&&ds^2=c^2\Bigl(\ds\f{\,3\,}{2}\sqrt{1-q a^2}-\f{\,1\,}{2}\sqrt{1-q
W^2}\Bigr)^2dt^2
-\nonumber \\[2mm]
&&\quad\;\;\,-(1-qW^2)^{-1}dW^2
 +W^2(d\theta^2+\sin^2\theta\,d\phi^2)\,;\label{eq57}
\ea
here $q=(1/3)\varkappa\rho =(2GM)/(c^2 a^3)$,\;$\varkappa = (8\pi
G)/c^2$,\;
$\rho =(3M)/(4\pi a^3)$.

\newpage 

The general property of external and internal solutions is manifested
in
the fact that at a definite value 
of $W$ the metric coefficients in front of the differential 
$dt^2$ in the intervals vanishes. 

Vanishing  of the metric coefficient 
$U$ of  $dt^2$ means that the gravitational field can, by its action,
not only slows down the lapse of time but can even 
 \textit{stop the flow of time}. For the external solution
 the vanishing  of the metric coefficient 
$U$ occurs at $W=W_g$. To exclude such a possibility (which is not
forbidden by the theory) one must assume that the body radius
satisfies the inequality
\be
a>W_g\,.
\label{eq58}
\ee
For the internal solution this happens at 
\be
W^2 = 9a^2 - 8(a^3/W_g)\, .
\label{eq59}
\ee
To exclude such a possibility of the vanishing of the metric
coefficient  $U$ inside the body one must assume that 
\be
a>(9/8)W_g\,.
\label{eq60}
\ee
\textit{One has to emphasize that inequalities  (\ref{eq58})  and 
(\ref{eq60}) 
are not a consequence of the GTR}.

The internal Schwazschild solution is somewhat formal but it is
interesting first all because it is an exact solution of the
equations of the GTR. In Part~2 it is shown taking as an example the
external Schwarzschild solution that in the relativistic theory of
gravitation, as in a field theory, inequality 
(\ref{eq58}) results exactly due to the stop of the process of
slowing down of the lapse of time. Below we consider an inertial
solution of the Schwarzschild type in the framework of the RTG.

The internal Schwarzschild solution arose on the basis of the
Hilbert--Einstein equations
\[
1-\ds\f{d}{dW}\Bigl[\f{W}{V}\Bigr]=\varkappa W^2\rho\,,
\]
\be
1-\f{1}{V}-\f{W}{UV}\f{dU}{dW}
=-\varkappa \f{W^2}{c^2}p\,.
\label{eq61}
\ee
Since, according to  (\ref{eq57}), the metric coeficients are
\be
U=\Bigl(\f{\,3\,}{2}\sqrt{1-q a^2}
-\f{\,1\,}{2}\sqrt{1-q W^2}\Bigr)^2,\quad
V=(1-q W^2)^{-1}, 
\label{eq62}
\ee
we find hence
\be
\f{\overset{\;\,\prime}{U}}{U}
=\f{q W}{\sqrt{1-q W^2}\Bigl(\ds\f{\,3\,}{2}\sqrt{1-q a^2}
-\ds\f{\,1\,}{2}\sqrt{1-q W^2}\Bigr)}\,,\quad
{\overset{\;\,\prime}{U}}=\f{dU}{dW}\,.
\label{eq63}
\ee
Substituting  (\ref{eq62}) and (\ref{eq63}) into (\ref{eq61}),
we obtain expression for the pressure 
\be
\f{p}{c^2}
=\f{\,\rho\,}{2}\ds\f{(\sqrt{1-q W^2}-\sqrt{1-q a^2})}{\sqrt{U}}\,.
\label{eq64}
\ee 
It is seen, in particular, from here that if equality 
(\ref{eq59}) were not excluded then the pressure inside the body on
the circumference, defined by this equality, would become infinite.
The singularity that arises due to the vanishing  of the metric
coefficient 
$U$ cannot be eliminated by the choice of the coordinate system
because the scalar curvature also has it:
\be
R=-8\pi G\Biggl[\f{3\sqrt{1-q a^2}-2\sqrt{1-q
W^2}}{\sqrt{U}}\Biggr]\,.
\label{eq65}
\ee
Let us show now, taking as an example an internal solution of the
Schwarzschild type, that the situation in the RTG changes drastically
due to the stop of the proces of the time lapse clowing-down.
The same mechanism of self-restriction which led, in the RTG, to 
inequality (\ref{eq58}) in the external Schwarzschild 
solution leads to an inequality of the type  (\ref{eq60}) 
for the internal Schwarzschild solution.

We will get the equations for this problem from 
(\ref{eq13'}) and (\ref{eq14'}). Introducing a new variable 
\[
Z=\f{UW^2}{V\,\acute{r}^2},\quad \acute{r}=\f{dr}{dW}
\]
and adding equations (\ref{eq13'}) and (\ref{eq14'}), we obtain 
\be
1-\f{1}{2UW}{\overset{\;\,\prime}{Z}}
+\f{m^2}{2}(W^2-r^2)
=\f{\,1\,}{2}\varkappa W^2\Bigl(\rho -\f{p}{c^2}\Bigr)\,.
\label{eq66}
\ee 

Subtracting equation  (\ref{eq14'}) from equation (\ref{eq13'}), we
find 
\be
{\overset{\;\,\prime}{Z}}-2Z\f{{\overset{\;\,\prime}{U}}}
{U}-2\f{Z}{W}-\f{m^2}{2}W^3\Bigl(1-\f{U}{V}\Bigr)
=-\varkappa W^3\Bigl(\rho +\f{p}{c^2}\Bigr)U\,.
\label{eq67}
\ee

In our problem the component of the energy-momentum tensor of the
substance are 
\[
T_0^0=\rho,\quad T_1^1=T_2^2=T_3^3=-\f{p(W)}{c^2}\,.
\]
The substance equation 
\[
\nabla_\nu (\sqrt{-g}\,T_\mu^\nu)
=\pa_\nu (\sqrt{-g}\,T_\mu^\nu)
+\f{\,1\,}{2}\sqrt{-g}\,T_{\sigma\nu}\pa_\mu g^{\sigma\nu}=0
\]
is reduced, for this problem, to the following form 
\be
\f{1}{c^2}\f{dp}{dW}=-\Bigl(\rho +\f{p}{c^2}\Bigr)
\f{1}{2U}\f{dU}{dW}\,.
\label{eq68}
\ee
Since the pressure increases towards the center of the ball this
leads  to an inequality 
\be
\f{dU}{dW}>0\,,
\label{eq69}
\ee
which indicates that with the  approach to the center of the ball
function 
$U$ decreases and hence a slowing down of the lapse of time occurs 
in
comparison with the inertial one. Since in the internal Schwarzschild
problem the density 
$\rho$ is assumed  \textit{constant},
equation (\ref{eq68}) is  solved  easily: 
\be
\rho+\f{p}{c^2}=\f{\alpha}{\sqrt{U}}\,.
\label{eq70}
\ee
Comparing (\ref{eq64}) and (\ref{eq70}), we find the constant
$\alpha$ 
\be
\alpha =\rho\sqrt{1-q a^2}\,.
\label{eq71}
\ee

Equations (\ref{eq66}) and (\ref{eq67}), on the  assumption  that 
\[
m^2(W^2-r^2)\ll 1,\quad (U/V)\ll 1\,,
\]
and after introduction of a new variable  $y=W^2$, take the form 
\be
{\overset{\;\,\prime}{Z}}=U(1-3qy)
+\f{\alpha\varkappa}{2}y\sqrt{U}\,,
\label{eq72}
\ee
\be
\sqrt{U}{\overset{\;\,\prime}{Z}}-\f{\,1\,}{y}Z\sqrt{U}
-4Z(\sqrt{U})^\prime
+\f{\alpha\varkappa}{2}y U
-\f{m^2}{4}y\sqrt{U}=0\,.
\label{eq73}
\ee
Here and further we use the notation 
${\overset{\;\,\prime}{Z}}=dZ/dy$.

In Part 2 in the analysis of the spherically symmetric Schwarzschild
solution we have seen that due to an effective gravitational force of
repulsion the metric coefficient 
$U$ defining the slowing down of the lapse of time,  as compared with
the
inertial one, does not vanish  even in a strong gravitational
field. 

That is why below we will investigate the behaviour of the solution
to
these equations in the region of small 
$y$. For a zero-mass graviton we get from expression 
(\ref{eq62})
for small $y$ 
\be
\sqrt{U}\simeq\f{\,1\,}{2}(3\sqrt{1-q a^2}-1)
+\f{qy}{4}+\f{1}{16}q^2y^2\,.
\label{eq74}
\ee
It is also seen from this expression that the function $\sqrt{U}$ 
for the internal Schwarzschild solution can be equal to zero if
\be
3\sqrt{1-q a^2}=1\,,
\label{eq75}
\ee
and this leads to an infinite value both of the pressure 
$\rho$ and the scalar density 
$R$. Since with the rest mass of the graviton equations 
 (\ref{eq72}, \ref{eq73}) stop the processes of slowing down of the
 lapse of time, 
  it is naturally to expect that equality 
 (\ref{eq75}) cannot take place in the  physical (real) 
range of the function  $\sqrt{U}$. On the basis of (\ref{eq74}) 
we will search for a solution to equations 
 (\ref{eq72}, \ref{eq73}) for the function  $\sqrt{U}$
in the form 
\be
\sqrt{U}=\beta
+\f{qy}{4}+\f{1}{16}q^2y^2\,,
\label{eq76}
\ee
where  $\beta$ is an unknown constant which has to be determined, 
making
use of Eqs. (\ref{eq72}, \ref{eq73}).

Substituting expression  (\ref{eq76}) into equation  (\ref{eq72})
and integrating,   we find
\be
Z=\beta^2 y+\f{y^2}{2}
\Bigl(\f{\beta q}{2}-3\beta^2 q+\f{\alpha\varkappa\beta}{2}\Bigr)
+\f{y^3}{3}\Bigl[\f{q^2}{8}\Bigl(\beta+\f{\,1\,}{2}\Bigr)
-\f{3\beta}{2}q^2+\f{\alpha\varkappa q}{8}\Bigl]\,.
\label{eq77}
\ee
Taking into account expressions  (\ref{eq76}) and (\ref{eq77}) in
equation (\ref{eq73}) and neglecting small terms of order 
$(my)^2$,  we obtain for the determination of the constant 
$\beta$ the equation 
\be
2\beta^2 q +\beta (q-\alpha\varkappa)+m^2/3=0\,.
\label{eq78}
\ee
To clarify the inference let us note that the  member containing
$y^2$ 
has the following form: 
\[
-\f{qy^2}{48}\bigl\{7[2\beta^2
q+\beta\,(q-\alpha\varkappa)]+3m^2\bigr\}\,.
\]
With account of equation  (\ref{eq78}) it can be reduced to 
\[
-\f{q}{72}m^2 y^2\,.
\]
Taking into consideration that by definition 
\[
\alpha\varkappa -q =\f{\varkappa\rho}{3}\bigl(3\sqrt{1-q
a^2}-1\bigr)\,,
\]
we find from equation (\ref{eq78}) 
\be
\beta =\f{3\sqrt{1-q a^2}-1+\Bigl[
\bigl(3\sqrt{1-q a^2}-1\bigr)^2
-(8m^2)/\varkappa\rho\Bigr]^{1/2}}{4}\,.
\label{eq79}
\ee
Thus the metric coefficient  $U$ defining the process of slowing down 
of the lapse of time as compared to the inertial on 
хода времени по сравнениі с инерциальнvм \textit{is not zero}.

If to put the graviton rest mass zero, expression 
(\ref{eq79}), as one could expect,  coincides exactly with the
constant terms of expression 
(\ref{eq74}). One can the minimum value of the quantity $\beta$  
from formula (\ref{eq65}): 
\be
\beta_{\min} =\Bigl(\f{m^2}{2\varkappa\rho}\Bigr)^{1/2}.
\label{eq80}
\ee

The quantity  $\beta$ in the function  $\sqrt{U}$ defines a bound for
the process of the lapse of time slowing down by the gravitational
field of the ball. It means that further 
\textit{slowing down} of the lapse of time by the gravitational field
is 
\textit{impossible}.
That is why the scalar cutvature defined by expression 
(\ref{eq65}) will be, in contrast to the GTR, finite everywhere. Thus
the very gravitational field stops, due to the rest mass  of the
graviton,  the process of the lapse of time slowing down.

According to  (\ref{eq79}) equality  (\ref{eq75}) is, due to the rest
mass of the graviton, 
\textit{impossible} because the inequality takes place 
\be
3\sqrt{1-q a^2}-1\geq
2\sqrt{2}\Bigl(\f{m^2}{\varkappa\rho}\Bigr)^{1/2}\,.
\label{eq81}
\ee
Since by definition the equality holds 
\[
q a^2=W_g/a\,,
\]
we find on the basis of inequality (\ref{eq81}) for $\varkappa\rho\gg
m^2$ 
\be
a\geq\f{\,9\,}{8}W_g\Biggl(1+\sqrt{\f{m^2}{2\varkappa\rho}}\,\Biggr)\
.
\label{eq82}
\ee
This \textit{limit for the radius of the body} arising from the study
of
the internal solution is 
\textit{stronger} than the limit   (\ref{eq58}) obtained 
in Part 2 from the analysis of the external solution. Inequality 
 (\ref{eq82}), as we see, follows directly from the theory, while in
 the GTR one has to specially introduce inequality 
 (\ref{eq60}) in order to avoid an infinite pressure  inside the
 body. On the basis of 
 (\ref{eq70}) and (\ref{eq71}) we find for the pressure:
\[
\f{p}{c^2}=\f{-\rho\sqrt{U}+\rho\sqrt{1-q a^2}}{\sqrt{U}}\,.
\]
Taking into account equality  (\ref{eq80}) 
we obtain the maximum pressure in the center of the ball 
\[
\f{p}{c^2}\simeq\rho\Bigl[\f{2\varkappa\rho}{m^2}(1-q
a^2)\Bigr]^{1/2}\,.
\]

The pressure in the center of the ball is finite, while in the GTR,
according to (\ref{eq57}), it is infinite. 

The self-restriction of the magnitude of the gravitational field 
arising in the relativistic theory of gravitation distinguishes it as
a matter of principle from the GTR and Newtonian theory of gravity in
which only 
 \textit{attractive forces} are present.  In the field theory of
 gravitation the presence of the rest-mass of the graviton and the
 fundamental property of the gravitational field to stop the process
 of the lapse of time slowing down imply that the 
\textit{gravitational force}  can be not only an 
\textit{attractive force} but at certain conditions (strong fields)
it manifests itself as an 
\textit{effective decelerating force}.  It is this force which stops
the proceses of a slowing down of the lapse of time by the
gravitational field. Thus the gravitational field cannot, in
principle, stop the lapse of time of a physical process because it
possesses a fundamental property of 
\textit{``self-restriction''}.

 In Parts~2, 3  we have seen that in
the GTR the metric coefficient~$U$ defining the slowing down of the
lapse of time by the gravitational field can become zero.
R.~Feynmann mentioned this circumstance and wrote on this
occasion~[7]: \textit{``\ldots if our formula for the time dilation
were correct then the physical processes would stop at the center of
the Universe because the time would not lapse there at all. This is
not
only physically unacceptable prediction: since we could expect that
the matter near the edge of the Universe would interact faster, the
light from the distant galaxies would have a violet shift. In fact it
is well known that it is shifted to lower, more red frequencies.
Thus, our formula for the time dilation is evidently needed to be
discussed in the following in connection with analysis of possible
models of the Universe. The following discussion is purely
qualitative and intended only to stimulate more wise thoughts on this
subject''.}

\section{On impossibility of the utmostly hard state
equation of~substance}

The self-restriction of the potential, as we have seen, is an
important property of the gravitational field. It is this property

that provides the presence of the bound of time dilation. 

Such a bound must exist because otherwise we come to a physically
inadmissible conclusion. Thus any metric field theory of the
gravitational field has to include this general statement as a 
 \textit{physical principle}.

The RTG combined with this general physical requirement allows to
ascertain that 
\textit{the utmostly hard state equation of the substance is not
realistic}. 

For the first time the question of the utmostly hard state equation
of the substance was discussed in the Ya.B.~Zeldovich paper~[8]. This
state
equation of the substance has the form:
\be
p/c^2=\rho -a\,,
\label{eq1}
\ee
here $p$ is the pressure, $\rho$ is the substance  density, 
$a$ is some constant. With such a state equation of the substance the
sound
velocity is equal to the velocity of light.

Let us consider, in the RTG, a spherically symmetric problem defined
by the interval 
\be
ds^2=c^2U(W)dt^2-V(W)dW^2
-W^2(d\theta^2+\sin^2\theta\,d\phi^2)
\label{eq2}
\ee
and the state equation of the substance
(\ref{eq1}). The state equation of the substance is written in the form
\vspace*{-1mm}

\be
\f{\,1\,}{c^2}\f{dp}{dW}=-\Bigl(\rho+\f{p}{c^2}\Bigr)\f{1}{2U}\f{dU}
{dW}\,.
\label{eq6}
\ee
Taking into account (83) we find from equation (85)
\vspace*{-1mm}

\be
\Bigl(\f{p}{c^2}+\f{\,a\,}{2}\Bigr)U=\alpha\,, 
\label{eq7}
\ee
here $\alpha$ is an integration constant.

The system of the RTG equations for interval (\ref{eq2}) 
in approximation (19) 
has the
form:
\be
{\overset{\;\,\prime}{Z}}-\f{2Z}{U}{\overset{\;\,\prime}{U}}
-2\f{Z}{W}=-2\Bigl(\varkappa\alpha -\f{\,m^2\,}{4}\Bigr)W^3\,,
\label{eq101}
\ee
\be
2UW-{\overset{\;\,\prime}{Z}}=\varkappa aW^3\,.
\label{eq102}
\ee
Here the function $Z$ is defined by the expression $Z=(UW^2)/V$.


This system of equations has an exact solution 
\be
U=2\Bigl(\varkappa\alpha -\f{\,m^2\,}{4}\Bigr)W^2,\;
Z=W^4\Bigl(\varkappa\alpha -\f{\,m^2\,}{4}\Bigr)
\Bigl(1-\f{\,1\,}{3}\varkappa a W^2\Bigr)\,.
\label{eq11}
\ee
From the definition of $Z$ we have 
\be
V=2\Bigl(1-\f{\,1\,}{3}\varkappa a W^2\Bigr)^{-1}\,.
\label{eq12}
\ee
At small values of $W$ Eq.~(15) 
is easily solved and one comes to the expression
\be
r={\mbox{const}}\, W^{\sqrt{5}-1}.
\label{eq13}
\ee
It is not difficult to see, making use of  (\ref{eq11}), (\ref{eq12})
and 
(\ref{eq13}), that at small  $W$ inequalities~(\ref{eq19})
hold rigorously. From
(88) and (\ref{eq11}) we find for the pressure, as a
\textit{scalar} quantity, the expression
\be
\f{p}{c^2}=-\f{\,a\,}{2}+
\f{\alpha}{2(\varkappa\alpha-m^2/4)W^2}\,.
\label{eq14}
\ee
That is why the singularity at  $W=0$ cannot be eliminated by the
choice of coordinate system. At 
$m^2=0$ solution (11) becomes the solution to the RTG, found
in
ref.~[9].

From expression  for $U$ it is evident that no
restriction for the time slowing down by the gravitational field
arises from the state equation of the substance   
(\ref{eq1}), and  therefore the pressure at the center, $W=0$,
according to~(\ref{eq14}), gets infinite, which is  physically
inadmissible. 

Thus, 
\textit{the utmostly hard state equation of the substance (83) does
not
realize} because it leads to the time stop at the center, 
$W=0$, and so 
\textit{violates} the above mentioned principle of the bound for the
time slowing down.

\section{ Is the Minkowski space observable?}

Now we ask a question: if the Minkowski space observable, at least in
principle?

To answer it we write down equations  (\ref{eq1}) in the form
\[
\f{m^2}{2}\gamma_{\mu\nu}
=8\pi G\Bigl(T_{\mu\nu}-\f{\,1\,}{\,2\,}g_{\mu\nu}T\Bigr)
-R_{\mu\nu}
+\f{m^2}{2}g_{\mu\nu}\,.
\]
It  is seen from here that in the r.h.s. there are only geometric
characteristics of the effective Riemannian space and quantities
which define the substance distribution in this space.

Now let us make use of the Weyl--Lorentz--Petrov theorem [10], 
according to which: ``\textit{If one knows \ldots the equations of
all
timelike
and all isotropic geodesic lines it is possible to determine the
metric tensor up to a constant multiplier}''. 
Hence it follows that with help of the experimental study of the
particles and the photon in the Riemann space one can, 
 \textit{in principle}, determine the 
\textit{metric tensor} $g_{\mu\nu}$ of the effective Riemannian
space. Substituting further $g_{\mu\nu}$ into the equation one can
determine the Minkowski space 
\textit{metric tensor}. After that one can , with help of coordinate
transformations, provide a passage to an 
\textit{inertial} Galilean coordinate system. Thus the Minkowski
space is, in principle, 
\textit{observable}.

It is proper to quote here the words by V.A.Fock [11]:
\textit{``How has one to define the straight line: as a light ray or
as a straight line in the Euclidean space in which harmonic
coordinates   $x_1,\,x_2,\,x_3$ are used as Cartesian coordinates? We
believe that the second definition is the only correct one. Actually
we used them when we said that the ray of light near the Sun has a
form of a hyperbola''}, and further  apropos of this: 
\textit{``\ldots consideration that the straight line, as a ray of
light, is more directly observable, it has no significance: what is
decisive in definitions is not their direct observability but rather
a correspondence to Nature, though this correspondence is established
by an indirect deduction''}.

The \textit{inertial} coordinate system, as we see, is related to
the substance destribution in the Universe. Thus, the RTG gives us, 
\textit{in principle},  an opportunity to determine the 
\textit{inertial} coordinate system. 

\section{ The evolution of homogeneous and isotropic Universe}
\subsection{The equations of the evolution of the scale factor}

In a homogeneous and isotropic Universe the interval in the effective
Riemannian space can be presented in the Friedmann--Robertson--Walker
metric:
\be
ds^2 =c^2U(t)dt^2
-V(t)\Bigl[\f{dr^2}{1-kr^2}+r^2 
(d\theta^2+\sin^2\theta\,d\phi^2)\Bigr]\,,
\label{eq107}
\ee
whereas the interval in the Minkowski space takes the form
\be
d\sigma^2 =c^2dt^2-dr^2 
-r^2(d\theta^2+\sin^2\theta\,d\phi^2)\,.
\label{eq108}
\ee
Let us write down equations   (1), (2)  of the RTG in
the form 
\be
\f{m^2}{2}\gamma_{\mu\nu}
=8\pi G\Bigl(T_{\mu\nu}-\f{\,1\,}{2}g_{\mu\nu}T\Bigr)
-R_{\mu\nu}+\f{m^2}{2}g_{\mu\nu}\,,
\label{eq109}
\ee
\be
\pa_\mu\tilde{g}^{\mu\nu}+
\gamma_{\lambda\sigma}^\nu\tilde{g}^{\lambda\sigma}=0\,.
\label{eq110}
\ee
With account of equations
\ba
&&\gamma_{22}^1=-r,\;\gamma_{33}^1=-r\sin^2\theta,\;
\gamma_{12}^2=\gamma_{13}^3=1/r,\nonumber \\[2mm]
&&\gamma_{33}^2=-\sin\theta \cos\theta,\;\gamma_{23}^3
=\ctg\theta,\nonumber \\[2mm]
&&\tilde g^{00} =V^{3/2}
U^{-1/2}(1-kr^2)^{-1/2}r^2\sin\theta,\nonumber \\[-2mm]
\label{eq111}\\[-2mm]
&&\tilde g^{11} = - V^{1/2} U^{1/2} (1-kr^2)^{1/2} r^2\sin\theta,
\nonumber\\[2mm] &&\tilde g^{22} = - V^{1/2} U^{1/2} (1-kr^2)^{-1/2}
\sin\theta,\nonumber \\[2mm]
&&\tilde g^{33} = - V^{1/2} U^{1/2} (1-kr^2)^{-1/2}
(\sin\theta)^{-1},\nonumber 
\ea
equations  (\ref{eq110}) for  $\nu =0$ and $\nu =1$ take the form 
\be
\f{d}{dt}\Bigl(\f{V}{U^{1/3}}\Bigr)=0\,,
\label{eq112}
\ee
\be
-\f{d}{dr}\Bigl[(1-kr^2)^{1/2}r^2\Bigr]+2(1-kr^2)^{-1/2}r=0\,.
\label{eq113}
\ee
For components  $\nu =2$ and $\nu =3$ equations  (\ref{eq110}) hold
identically. It folows from equations 
 (\ref{eq112}) and  (\ref{eq113}) that
\be
V/U^{1/3}=\mbox{const}=\beta^4\ne 0,\quad k=0\,.
\label{eq114}
\ee
Thus, since the system of equations of the RTG is complete, it 
\textit{leads unambiguously, in contrast to the GTR, to a unique
solution, that is the flat spatial (Euclidean) geometry of the
Universe. }

Assuming that
\be
a^2=U^{1/3},
\label{eq115}
\ee
we obtain
\be
ds^2 =\beta^6\Bigl[c^2d\tau_g^2-\Bigl(\f{\,a\,}{\beta}\Bigr)^2 
(dr^2+r^2d\theta^2+r^2\sin^2\theta\,d\phi^2)\Bigr]\,.
\label{eq116}
\ee
Here the quantity 
\be
d\tau_g=\Bigl(\f{\,a\,}{\beta}\Bigr)^3dt
\label{eq117}
\ee
determines the rate of dilating of the lapse of time in the presence
of the gravitational field in comparison with the inertial time 
$t$.

Common constant numerical factor  $\beta^6$ in the interval 
$ds^2$ equally increases both the time and the spatial variables. It
does not reflect  the dynamics of the Universe development but
determines the time of the Universe and its spatial scale. 
The time of the Universe is determined by the quantity 
$d\tau$ as a timelike part of the interval 
$ds^2$ 
\be
d\tau =\beta^3d\tau_g=a^3dt,
\label{eq118}
\ee
\be
ds^2 =c^2d\tau^2-\beta^4a^2(\tau) 
(dr^2+r^2d\theta^2+r^2\sin^2\theta\,d\phi^2)\,.
\label{eq119}
\ee

The energy-momentum tensor of the substance in the effective
Riemannian space takes the form 
\be
T_{\mu\nu}=(\rho +p)U_\mu U_\nu -g_{\mu\nu}p\,,
\label{eq120}
\ee
where $\rho$ and $p$ are the density and the pressure, respectively,
of the substance in its rest frame,   while 
$U_\mu$  is its velocity. Since $g_{0i}$ and  $R_{0i}$
are zero for interval (\ref{eq119}), it follows from equation   
(\ref{eq109}) that 
\be
T_{0i}=0\;\;\mbox{и}\;\;U_i=0\,.
\label{eq121}
\ee
This means that in the  \textit{inertial frame} 
defined by the interval  (\ref{eq108}) the 
\textit{substance} during the Universe evlution  remains in the state
of rest. Immobility of the substance in the homogeneous and
isotropic Universe (leaving aside peculiar velocities of Galaxies) 
corresponds, in some sense, to early (pre-Friedmann) A. Einstein's 
concepts of the Universe. 

So-called ``expansion of the Universe'', observed with the red shift, 
is caused 
\textit{not by the substance motion} but rather by 
 \textit{the change} of the  \textit{gravitational field}  with time.
 One has to keep in mind this observation when using the established
 term ``expansion of the Universe''.

With the description of interval  (\ref{eq119}) in the proper time
$\tau$
the interval of the primordial Minkowski space (\ref{eq108}) assumes
the form
\be
d\sigma^2=\f{c^2}{a^6}d\tau^2-dr^2-r^2(d\theta^2
+\sin^2\theta\,d\phi^2)\,. 
\label{eq122}
\ee
On the basis of (\ref{eq119}) and (\ref{eq122}) and taking into
account
that
\be
R_{00}=-3\f{\ddot{\,a\,}}{a},\quad
R_{11}=\beta^4(a\ddot{a}+2\dot{a}^2)\,,
\label{eq123}
\ee
\be
T_{00}-\f{\,1\,}{2}g_{00}T=\f{\,1\,}{2}(\rho +3p),\quad
T_{11}-\f{\,1\,}{2}g_{11}T=\f{\,1\,}{2}\beta^4 a^2(\rho -p)\,,
\label{eq124}
\ee
we obtain from equations (\ref{eq109}) the equations for   the scale
factor 
\be
\f{\,1\,}{a}\f{d^2 a}{d\tau^2}
=-\f{4\pi G}{3}\Bigl(\rho + \f{3p}{c^2}\Bigr)
-\f{\,1\,}{6}(mc)^2\Bigl(1-\f{1}{a^6}\Bigr)\,,
\label{eq125}
\ee
\be
\Bigl(\f{\,1\,}{a}\f{da}{d\tau}\Bigr)^2
=\f{8\pi G}{3}\rho (\tau)-
\f{1}{12}(mc)^2 \Bigl(2-\f{3}{a^2\beta^4}+\f{1}{a^6}\Bigr) \ .
\label{eq126}
\ee

In the absence of the substance and gravitational waves equations 
(\ref{eq125}), (\ref{eq126}) have a trivial solution 
$a=\beta=1$, i.e. the evolution of the empty Universe does not occur
and the effective Riemannian space coincides with the Minkowski
space.
 \textit{Let us note that in the developed theory the absolute
 meaning
 of the scale factor $a$ acquires the physical sense}.
At  $m=0$ equations (\ref{eq125}) and (\ref{eq126})
coincide with the Friedmann equation for the evolution of the flat
Universe. However, the presence of terms with 
 $m\ne 0$ essentially changes the character of evolution at small and
 large values of the scale factor. 

The appearance of additional terms in equations 
 (\ref{eq125}) and (\ref{eq126}) at 
$m^2\ne0$ (in particular, of the terms 
$\sim m^2/a^6$)
is related to the passage from the inertial system time $t$ to the
time 
$\tau$ (\ref{eq118}). Since the gravitation influences the lapse of
time, these terms appear to be large enough to influence the
character of evolution in the strong gravitational fields (in spite
of the smallness of the graviton mass).

Specifically, because of this change of the lapse of time in the
gravitational field the forces arise that are manifested as
repulsive forces during the shrinking of the Universe or as
attractive forces in the final stage of the expansion.

The proportionality of the r.h.s. terms in
 (\ref{eq125}) and (\ref{eq126})
to the square of the rest mass of the graviton is a manifestation of
the fact that only at 
$m^2\ne0$ the effective Riemannian space preserves the connection
with the basic Minkowski space.

\subsection{Absence of the cosmological singularity}

From the covariant conservation law of the energy-momentum tensor
$\tilde T^{\mu\nu}=\sqrt{-g}\,T^{\mu\nu}$
$$
\nabla_\mu \tilde T^{\mu\nu}
=\partial_\mu \tilde T^{\mu\nu}
+\varGamma^\nu_{\alpha\beta}\tilde T^{\alpha\beta}=0 \ , 
$$
where  $\nabla_\mu$ is the covariant derivative and 
$\varGamma^\nu_{\alpha\beta}$  are the Christoffel symbols in the
Riemannian space, which follows from 
(1), (2)  and from expression  (\ref{eq120}), the expression results 
\be
-\f{\,1\,}{a}\f{da}{d\tau}
=\f{1}{3\Bigl(\rho+\ds\f{p}{c^2}\Bigr)}\f{d\rho}{d\tau}\,. 
\label{eq127}
\ee
For the state equation of the substance $p=f(\rho)$ 
equation (\ref{eq127}) determines the dependence of the substance
density on the scale factor. In the case when the state  equation has
the form 
\[
\f{p}{c^2}=\omega\rho\,,
\]
this dependence is given by the expression 
\[
\rho=\f{\mbox{const}}{a^{3(\omega+1)}}\,.
\]
For the cold substance, including the dark and baryon masses, 
$\omega_{CDM}=0$; for the radiation density  $\omega_r=1/3$, and
for the quintessence
$\omega_q=-1+\nu$. Thus the total substance density in Eqs.
(\ref{eq125}) and (\ref{eq126}) has the form:
\begin{equation}
\rho=\f{A_{CDM}}{a^3} +\f{A_r}{a^4}+\f{A_q}{a^{3\nu}}\,,
\label{eq128}
\end{equation}
where $A_{CDM}$, $A_r$ and $A_q$  are constant quantities. According
to  (\ref{eq128}) the radiation dominated stage of the Universe
evolution takes place at small values of the scale parameter 
\mbox{$(a\ll 1)$}: 
\[
\rho\approx \rho_r=\f{A_r}{a^4}\,.
\]
Turning to equation  (\ref{eq126}), one can notice that at  $a\ll 1$
the negative term in the r.h.s. of the equation grows with
decreasing of the scale factor as 
$1/a^6$ in modulus. Since the l.h.s. of the equation is positive
definite there must exist a minimum value of the scale factor 
\be
a_{\min}=\f{mc}{(32\pi G A_r)^{1/2}}=
\left ( \f{m^2c^2}{32\pi G \rho_{\max}}\right )^{\!\!1/6}\!\!. 
\label{eq129}
\ee
The existence of the minimal value of the factor  (\ref{eq129})
means that the process of slowing down of the lapse of time by the
gravitational field during the compression of the Universe stops.
Therefore, the gravitational field cannot stop the lapse of time.

Thus, \textit{due to  the graviton mass and, hence, to the  
presence of the gravitational forces related with the change of the
lapse of time the cosmological singularity is eliminated},
and the expansion of the Universe starts 
from a finite value of the scale factor 
(\ref{eq129}). Specifically, here a surprising property of the
gravitational field is manifested: an ability to create in the strong
fields the
repulsive forces which stop the process of the compression of the 
Universe and then provide its accelerated expansion.

On the basis of  (\ref{eq125}) and 
(\ref{eq129}) we determine the initial acceleration which was a
\textit{``push''} to the expansion of the Universe. 

It is 
\[
\f{\,1\,}{a}\ds\f{d^2 a}{d\tau^2}\bigg|_{\tau =\,0}
=\f{8\pi G}{3}\rho_{\max}\,, 
\]
and hence, in the RTG, in the radiation dominated stage of the
Universe 
\textit{in the period of the accelerated expansion}, which
\textit{precedes}
the Friedmann expansion stage, the scalar curvature is not zero and
at 
$\tau =0$ 
\[
R=-\f{16\pi G}{c^2}\rho_{\max}\,,
\]
while it is equal to zero in the GTR. When the scale factor
$a(\tau)$
is 
\[
a^2(\tau)=\f{\,3\,}{2}a_{\min}^2\,,
\]
the Habble constant gets maximum
\[
H_{\max}=3^{-2}(32\pi G\rho_{\max})^{1/2}\,,
\]
the scalar curvature  $R$ is 
\[
R=-\Bigl(\f{\,2\,}{3}\Bigr)^3\f{16\pi G\rho_{max}}{c^2}\,,
\]
and
\[
R_{\rho\lambda\mu\nu}R^{\rho\lambda\mu\nu}=8\cdot 3^{-7}
\Bigl(\f{32\pi G}{c^2}\rho_{\max}\Bigr)^2.
\]
Since the scalar curvature 
 $R$ and the invariant 
$R_{\rho\lambda\mu\nu}R^{\rho\lambda\mu\nu}$
depend of $\rho_{max}$ one can expect an intensive production of the
gravitons in the radiation dominated stage. 

In such a way a relativistic relic gravitational background of
non-thermal  origin can arise. 

\subsection{Impossibility of the unlimited ``expansion of the
Universe''}

Considering the gravitational field  $\phi^{\mu\nu}$ as a physical
field in the Minkowski space, one has to require the fulfilment of
the causality principle. This means that the light cone in the
effective Riemannian space has to lie inside the light cone of the
Minkowski space, i.e. for 
$ds^2=0$ the requirement 
$d\sigma^2\geqslant0$ holds. Writing down  $d\sigma^2$ in the
spherical 
coordinate system
\begin{equation}
d\sigma^2=c^2dt^2-(dr^2+r^2d\theta^2+r^2\sin^2\theta\,d\phi^2) 
\label{eq130}
\end{equation}
and determining the spatial part of the interval from the condition 
$ds^2=0$, we have
\[
d\sigma^2=c^2dt^2 \left (
1-\f{a^4}{\beta^4} \right )\geqslant0,
\]
i.e.
\begin{equation}
(a^4-\beta^4)\leqslant0.
\label{eq131}
\end{equation}
Thus the scale factor  $(a)$ is bounded by the condition 
$a\leqslant\beta$ and 
so it would be natural to assume its maximum value as 
\[
a_{\max}=\beta\,. 
\]
With such a choice of  $a_{\max}$ the rate of the  lapse of time
$d\tau_g$
in the moment  of the stop of the Universe expansion becomes equal to
the rate of the lapse of the inertial time 
 $t$  in the Minkowski space, though the second derivative 
 ${\ddot a}$ and, hence, the scalar curvature 
 $R$ are non-zero. This is this point from which the slowing down of
 the rate of the lapse of time under the action of the attractive
 forces will proceed up to the point of the stop of compression, when
 under the action now already of repulsive forces the opposite
 process of the acceleration of the rate of the lapse of time up to
 the rate of the inertial time $t$ of the Minkowski space starts.
\textit{ Exactly
 all these physical consequences require necessarily the condition 
$a_{max}=\beta$ to be held. }
As we will see further (see Part. 6.7), the value of the quantity  
 $\beta$ is determined by the integral of motion. 

Condition (\ref{eq131}) does not admit an inlimited growth of the
scale factor with time $\tau$, i.e. an unlimited 
``expansion'' of the Universe (in the above-indicated sense) which is
provided by the dynamical evolution equation of the scale factor $a$.
Let us note, besides, that  the very Universe is infinite because the
radial coordinate is defined in the range 
$0<r\leq\infty$.

\subsection{Evolution of the early Universe}

In the radiation dominated stage of the Universe $(\rho=\rho_r)$ at
\mbox{$a\ll 1$} equations (\ref{eq125}), (\ref{eq126}) assume the
form
\be
\left (
\f{\,1\,}{\xi}\f{d\xi}{d\tau}\right )^2
=\f{1}{\tau^2_r}
\left (
1-\f{1}{\xi^2}\right )
\f{1}{\xi^4}\ , 
\label{eq132}
\ee
\be
\f{\,1\,}{\xi}\f{d^2\xi}{d\tau^2}
=\f{1}{\tau^2_r}
\left (
\f{2}{\xi^2}- 1\right )
\f{1}{\xi^4}\ ,
\label{eq133}
\ee
where
\[
\xi=\f{a(\tau)}{a_{\min}};\;\;
\tau_r =\left (
\f{3}{8\pi G\rho_{\max}}
\right )^{\!\!1/2}\!\!.
\]
The solution to equation (\ref{eq132}) is
\begin{equation}
\f{\tau}{\tau_r}=\f{\,1\,}{2}
\left\{
\xi (\xi^2-1)^{1/2}
+\ln [\xi+(\xi^2-1)^{1/2}]\right\}\,.
\label{eq134}
\end{equation}
It  $\xi-1\ll 1$ $(\tau\ll\tau_r)$:
\[
a\simeq a_{\min}
\left \{
1+\f{\,1\,}{2}
\left (
\f{\tau}{\tau_r}
\right )^2
-\f{7}{24}
\left (
\f{\tau}{\tau_r}\right )^4
\right\}.
\]

Adding equations  (\ref{eq132}) and (\ref{eq133}), we obtain
\[
\ddot{a}/a+(\dot{a}/a)^2=(mc)^2/12a^6\,,\mbox{где}\;\dot{a}=da/d\tau\
.
\]
In the GRT the l.h.s. of this equation in the radiation dominated
domain is equal to zero exactly and hence the Friedmann stage takes
place when the scale factor 
$a(\tau)$ changes with time as  $\tau^{1/2}$. In the RTG, according
to this equation, there exists, in the radiation dominated phase, 
``pre-Friedmann'' stage of the Universe evolution where the scalar
curvature 
 $R$ is 
\[
R=-\f{\,1\,}{2}(mc)^2\f{\,1\,}{a^6}\,.
\]

Here the particle horizon is 
\[
R_{\mbox{part}}(\tau)=a(\tau)
\int\limits^{\tau}_{0}\!\!\f{c\,d\tau'}{a(\tau')}
\simeq c\tau
\left (
1+\f{\,1\,}{3}\f{\tau^2}{\tau^2_r}\right ).
\]
The accelerated expansion proceeds, according to (\ref{eq133}), till
the values 
$\xi=\sqrt 2$ (i.\,e. $a=\sqrt 2\,a_{\min}$) during the time 
\[
\tau_{in}=\tau_r
\f{\,1\,}{2}
(\s 2 +\ln (1+\s 2))
\simeq 1.15\tau_r\ .
\]
The quantity  ${\dot a}/a$ achieves its maximum value 
$[{\dot a}/a]_{max} = 2/3\sqrt{3} \tau_{r}$ somewhat earlier, at 
$a/a_{min}=
\sqrt{3/2}$ and at $\tau\sim 0{,}762\,\tau_{r}$. \textit{The large
acceleration during the growth of the scale factor starting from its
minimum value 
$({\ddot a}/a)_{0}= 1/\tau_{r}^{2}$ is related with effective forces
arising because of the difference in the lapse of time 
 $t$ and $\tau$ (see equation (\ref{eq118})),
caused by the action of gravitation}. 
Exactly
these forces lead to the terms 
$~m^{2}/a^{6}$ in equations (\ref{eq125}), (\ref{eq126}).
At  $\tau > \tau_{in}$ the acceleration changes to the deceleration.
At $\xi\gg 1$ the expansion (\ref{eq134}) passes to the Friedmann
regime corresponding to the radiation dominated stage 
\[
a_{(\tau)}=a_{\min}\,\xi
\simeq a_{\min} \Bigl(\f{2\tau}{\tau_r}\Bigr)^{\!1/2}\!, 
\]
at to the dependence known for this regime 
\begin{equation}
\rho \simeq \rho_r (\tau)
=\f{3}{32\pi G\tau^2};\;\;\; \tau \gg \tau_r\ .
\label{eq135}
\end{equation}

In order that in the first seconds since the beginning of the
expansion the laws of primordial nucleosynthesis to be held it is
necessary that 
$\tau_r\lesssim 10^{-2}\mbox{[s]}$. The bound for 
 $\rho_{\max}$ corresponding to this requirement is rather weak:
\[
\rho_{\max}> 2\cdot 10^{10}~\mbox{[g}\cdot \mbox{cm}^{-3}\mbox{]}\,.
\]
The value $\rho_{\max}$ at energies $kT\simeq 1\,\mbox{TeV}$ 
corresponding to the electroweak scale and with account of all
degrees of freedom of leptons, quarks etc. is
\[
\rho_{\max}\simeq 10^{31}~\mbox{[g}\cdot \mbox{cm}^{-3}\mbox{]}\,,
\]
while at the Grand Unification scale  $kT\simeq 10^{15}\,\mbox{GeV}$
\[
\rho_{\max}\simeq 10^{79}~\mbox{[g}\cdot \mbox{cm}^{-3}\mbox{]}\,.
\]

Thus, since the scale factor cannot become zero, this means that
according to the RTG no  
 \textit{Big Bang} could occur in the Universe. 
In the past, everywhere in the Universe, {\it the substance} was in the
gravitational field in the state of high density and high
temperature, and then it evolved as was described above.

\subsection{Total relative density of the substance and the graviton
mass}

Let $a_0$ is the present value of the scale factor, while 
$\rho^0_c$ is the critical density related to the present value of
the
Hubble constant 
$H=\Bigl(\ds\f{\,1\,}{a}\f{da}{dt}\Bigr)_0$ by the relation 
\[
H^2=\f{8\pi G}{3}\rho^0_c\ .
\]
Introducing the variable 
\[
x=\f{a}{a_0}\ ,
\]
and the ratios of the densities 
\[
\Omega^0_r=\f{\rho^0_r}{\rho_c^0};\;\;\;
\Omega^0_m=\f{\rho^0_m}{\rho_c^0};\;\;\;
\Omega^0_q=\f{\rho^0_q}{\rho_c^0}\,, 
\]
one can, with account of relation  (\ref{eq128}), write down
equations 
(\ref{eq125}), (\ref{eq126}) in the form 
\be
\left (\f{\,1\,}{x}\f{dx}{d\tau}\right )^2
= H^2 \Biggl \{
\f{\Omega^0_r}{x^4}
+\f{\Omega^0_m}{x^3}
+\f{\Omega^0_q}{x^{3\nu}}
-\f{f^2}{6}
\left (
1-\f{3}{2\beta^4 a^2}
+\f{1}{2a^6}\right ) \Biggr \};
\label{eq136}
\ee
\be
\left (\f{\,1\,}{x}\f{d^2x}{d\tau^2}\right )
=
-\f{H^2}{2}\Biggl \{
\f{2\Omega^0_r}{x^4}
+\f{\Omega^0_m}{x^3}
-2 \left (1-\f{3\nu}{2}\right )
\f{\Omega^0_q}{x^{3\nu}}+ \f{f^2}{3}
\left (
1-\f{1}{a^6}\right ) \Biggr \},
\label{eq137}
\ee
where
\begin{equation}
f=\f{mc}{H}=\f{m_gc^2}{\hbar H}\ .
\label{eq138}
\end{equation}
For the present value of the quantities at $a_0\gg 1$ equation
(\ref{eq136}) gives the relationship
\[
1=\Omega^0_{\mbox{tot}}
-\f{f^2}{6},
\]
i.e. the total relative density is equal to 
\begin{equation}
\Omega^0_{\mbox{tot}}=\f{\rho^0_{\mbox{tot}}}{\rho^0_c}
=\Omega^0_r +\Omega^0_m+\Omega^0_q=1+\f{f^2}{6}\ .
\label{eq139}
\end{equation}

Hence the University possessing (according to the RTG) the Euclidean
spatial geometry has to have 
 $\Omega^0_{\mbox{tot}}>1$, while in the theories with a primordial
 inflationary expansion leading to the flat geometry the condition 
$\Omega^0_{\mbox{tot}}=1$ has to be satisfied with a great accuracy 
$(\sim 10^{-3}\div 10^{-5})$. Equation 
(\ref{eq139}) gives a possibility to evaluate the graviton mass from
the recent  experimental measurements of 
 $\Omega^0_{\mbox{tot}}$ and $H$.

\subsection{Upper limit for the graviton mass}

Determining the cosmological parameters from the observation of the
angular asymmetry of the micro-wave background radiation 
(or CMB)~[12] leads  systematically to an average value 
$\Omega^0_{\mbox{tot}}>1$.  This concerns both the first quantitative
experimental data from 
COBE~[13],  \mbox{Maxima-1~[14]} и Boomerang-98~[15] joint data processing
of which[16] gives the value  $\Omega^0_{\mbox{tot}}=1{,}11\pm
0{,}07$, 
and excellent data of the experiment WMAP~[17] which alone (without
taking into account 
the data on observation of the supernovae  SNIa [12, 18] 
and the galaxy catalogue (2dFGRS~[19] and SDSS~[20]))  give,
dependent on the 
choice of parameters, the values 
$\Omega^0_{\mbox{tot}}\!\!\!=\!\!\!1{,}095^{+0{,}094}_{-0{,}144}$
and 
$\Omega^0_{\mbox{tot}}\!=\!1{,}086^{+0,057}_{-0{,}128}$~[17]. 

Within the error bars these values certainly do not contradict the 
value $\Omega^0_{\mbox{tot}}=1$ following from the inflationary
model, but they can  indicate also the existence of the non-zero
graviton mass, according to relations 
(\ref{eq138}), (\ref{eq139}).
At any rate if to take the value 
 $\Omega^0_{\mbox{tot}}=1{,}3$
 which exceeds more than $2\sigma$ the average value of 
 $\Omega^0_{\mbox{tot}}$ then we obtain from 
(\ref{eq138}), (\ref{eq139}) with probability 95\% the upper bound
for the graviton mass. It is convenient to represent the quantity $f$
from (\ref{eq138}) in the form of the ratio of the graviton mass to
the quantity
\[
m_H=\f{\hbar H}{c^2}=3{,}80\cdot 10^{-66}\,h \ ,
\]
which could be called the  ``Hubble mass''. At  $f^2/6=0{,}3$
the upper limit for the graviton mass is 
\[
m_g\leqslant 1{,}34\,m_H\approx 5{,}1\cdot 10^{-66}\,h~\mbox{[g]}\,,
\]
or, at $h=0{,}70$,
\begin{equation}
m_g< 3{,}6\cdot 10^{-66}~\mbox{[g]}\,. 
\label{eq140}
\end{equation}
The Compton wavelength of the graviton appears to be compatible with
the Hubble radius of the Universe, $c/H$
\[
\f{\hbar}{m_g c}\lesssim 0{,}75\,\f{c}{H}\ . 
\]

The estimates of the upper limit for the graviton mass obtained
earlier were based on the fact that the gravitational potential with
non-zero graviton mass has to have the Yahawa form. 

On the basis of analysis of the dynamics of the galaxy clusters and
conservative estimates of the distances 
 $(\sim 580$ kps) at which the gravitational connection between the
 galaxies in the cluster still holds there was obtained in 
 works [21, 22] un upper limit for the graviton mass
\[
m_g < 2\cdot 10^{-62}~\mbox{[g]}\,. 
\]
Our estimate  (\ref{eq140}) improves this bound more than  5000
times. This is related to the fact that a consistent consideration of
the gravitational field in the Minkowski space includes not only the
equation, according to which the potential of the weak gravitational
field has the Yukawa form, but also general equations of gravitation  
 (\ref{eq1'}), (\ref{eq2'}), which are in agreement with all
 gravitational phenomenae in the Sun system and are applicable to the
 whole Universe, i.e. at distances of the order 
 $c/H\simeq 10^{28}\mbox{[cm]}$, more than  5000 times longer than
 the distances among gravitationally binded galaxies in the clusters.

\subsection{Integral of the Universe evolution and present value of
the scale factor}

Making use of relation (\ref{eq127}) one can exclude the pressure
from 
equation 
(\ref{eq125}), bring it to the form 
\[
\f{\,1\,}{a}\f{d^2a}{d\tau^2}
=\f{4\pi G}{3}\Bigl(a\f{d\rho}{da}+2\rho\Bigr)
-\f{\,1\,}{6}(mc)^2\Bigl(1-\f{1}{a^6}\Bigr)\,,
\]
and to write it further in the form
\be
\f{d^2a}{d\tau^2}+\f{dV}{da}=0\,,
\label{eq141}
\ee
where 
\be
V=-\f{4\pi G}{3}a^2\rho 
+\f{(mc)^2}{12}\Bigl(a^2+\f{1}{2a^4}\Bigr)\,.
\label{eq142}
\ee
Multiplying both sides of equation (\ref{eq141}) by 
$\ds\f{da}{d\tau}$, we obtain 
\[
\f{d}{d\tau}\Bigl[\f{\,1\,}{2}\Bigl(\f{da}{d\tau}\Bigr)^2+V\Bigr]=0\,
,
\]
or 
\be
\f{\,1\,}{2}\Bigl(\f{da}{d\tau}\Bigr)^2+V=E=\mbox{const.}
\label{eq143}
\ee
Expression  (\ref{eq143})
reminds the energy of the unit mass. If the quantity $a$ had the
dimension of length then the first term in
(\ref{eq143}) would correspond to the kinetic energy and the second
to the potential one. The quantity 
 $\Bigl(-\ds\f{4\pi G}{3}\rho a^2\Bigr)$
in (\ref{eq142}) corresponds to the gravitational potential on the
boundary of the ball of radius 
$a$ filled with a substance of constant density 
$\rho$ while the extra terms in 
(\ref{eq142})proportional to  $m^2$ 
correspond to effective forces arising, as was mentioned above,
because of the influence of gravitation on the lapse of time. 

The quantity  $E$ is the integral of the Universe evolution. It is
extremely small but at  $m\ne 0$ is not zero. Having substituted 
 $(da/d\tau)^2$ from equation  (\ref{eq126}) 
into equality  (\ref{eq143}), we get
\be
E=\f{(mc)^2}{8\beta^4}\,.
\label{eq144}
\ee
In such a way the constant  $\beta$ (see (\ref{eq131})) entering
interval 
(\ref{eq119}) and, according to (\ref{eq131}), limiting the growth of
the scale factor  
$a$
is expressed via the integral of motion $E$.

In what follows we shall need the present value of the scale factor, 
$a_0$.
One can obtain an estimate of this quantity from the following
considerations. Assuming that the Universe evolution begins in the
radiation dominated epoch, we have for the ratio 
$a_0/a_{\min}$ 
\[
\f{a_0}{a_{\min}}=\Bigl(\f{\rho_{\max}}{\rho_r^0}\Bigr)^{\!1/4}\!,
\]
where  $\rho_r^0$  is the present density of the radiation energy.  
In its turn 
$\rho_r^0$ can be expressed in terms of the relative density 
$\Omega_r^0$ and critical density  $\rho_c^0$
\[
\rho_r^0
=\Omega_r^0\rho_c^0
=\Omega_r^0\Bigl(\f{3H^2}{8\pi G}\Bigr)\,.
\]
Thus
\[
\f{a_0}{a_{\min}}=\Bigl(\f{8\pi}{3}\f{G\rho_{\max}}{H^2\Omega_r^0}
\Bigr)^{1/4}
\approx 1,34\cdot 10^{10}(G\rho_{\max})^{1/4}\,,
\]
where  $G\rho_{\max}$ is taken in  $\mbox{sec}^{-2}$. (In the course
of the calculation of the numerical factor in the above-said
expression we used the standard value 
$H=h/3,0857\cdot 10^{17}c$ and $\Omega_r^0=\Omega_\gamma^0=2,471\cdot
10^{-5}/h^2$).

Then, making use of definition  (\ref{eq138}), one can represent the
value $a_{\min}$ from  (\ref{eq129}) in the form 
\[
a_{\min}=\Bigl(\f{f^2}{6}\Bigr)^{1/6}
\Bigl(\f{3}{16\pi}\f{H^2}{G\rho_{\max}}\Bigr)^{1/6}
=8,21\cdot
10^{-7}\Bigl(\f{f^2}{6}\Bigr)^{1/6}\f{1}{(G\rho_{\max})^{1/6}}\,,
\]
where, according  (\ref{eq139})
\[
\f{f^2}{6}=\Omega_{\tot}^0-1\,,
\]
$a_{\min}$ at the electroweak scale is 
\[
a_{\min}\simeq 5\cdot10^{-11}\,,
\]
and at the scale of the Grand Unification 
\[
a_{\min}\simeq 5\cdot10^{-19}\,.
\]
For the quantity $a_0$ we have from the ratio $a_0/a_{\min}$
\footnote{In the course of the numerical estimate one takes as a
relative
density of the relativistic particles $\Omega_r^0$ the relative
density of the microwave relativistic radiation 
$\Omega_\gamma^0$ because it follows from the data on neutrino
oscillations that  at least two types of neutrino are nonrelativistic
at present time. When extrapolating to the early Universe one should
certainly take into account that the temperature of the relic
radiation in the course of the evolution raised due to 
$e^{+}e^{-}$-annihilation; before the onset of the annihilation it
was equal to the temperature of the neutrino gaz which at that time
also consisted of relativistic neutrinos and contributed into the
total density of the relativistic particles. In the same way, when
extrapolating to the early Universe, the density of the relativistic
gaz raises due to relativisation of other particles that are
produced. However, due to the fact that the quantity 
$\Omega_r^0$  enters (\ref{eq145}) in the form of 
$(\Omega_r^0)^{1/4}$, the numerical estimate  (\ref{eq145})
changes no more than three times (even if to assume that the number
of
degrees of freedom in the relativistic gaz is about 
100).}
\be
a_0=\Bigl(\f{f^2}{6}\Bigr)^{1/6}\Bigl(\f{2\pi}{3}\f{G\rho_{max}}{H^2}
\Bigr)^{1/12}
\f{1}{(\Omega_r^0)^{1/4}}\simeq 1,1\cdot 10^4
\Bigl(\f{f^2}{6}\Bigr)^{1/6}
(G\rho_{max})^{1/12}\,,\;\,
\label{eq145}
\ee
where $a_0$ at $\rho_{\max}$ chosen at the electroweak scale is 
\[
a_0\simeq 5\cdot10^5\,,
\]
and at the scale of the Grand Unification
\[
a_0\simeq 5{,}5\cdot10^9\,.
\]
As was already mentioned (see Part  6.1.) in the RTG an absolute 
value of the scale factor acquires a sense. At the average value 
$\Omega_{\tot}=1{,}02$
(i.e. $f^2/6=0{,}02$) and 
$\rho_{\max}$
$\gtrsim 10^{10}\mbox{[g}\!\cdot\!\mbox{cm}^{-3}\mbox{]}$
the quantity  $a_0\gg1$. This justifies the approximations made
during the inference of equality~(\ref{eq139}).

\subsection{Incompatibility of the RTG with the existence of a
constant cosmological term 
({\mathversion{bold}$\Lambda$}TDі theory). 
Necessity of quintessence with~{\mathversion{bold}$\nu>0$}}

As was already mentioned, when considering the gravitational field as
a physical field in the Minkowski space, it is necessary to require
the fulfilment of the causality principle. This requirement, applied
to the Universe evolution, leads to inequality 
(\ref{eq131}) according to which the scale factor is bounded by the
inequality
 $a\leqslant a_{\max}=\beta$. In other words, according to the RTG, 
 the unlimited expansion of the Universe is impossible. The
 mathematical apparatus of the RTG automatically provides the
 fulfilment of this condition in the case
\textit{when the matter density decreases with increase of the scale
factor}.
In fact, the structure of the term proportional to  $m^2_g$ in
equation (\ref{eq126}) is such that due to positive definitness of
the l.h.s. of the equation the third term in the bracket provides the
absence of the cosmological singularity at  \mbox{$a \ll 1$} while
the first term limits the minimum value of the matter density (and by
this limits from above the value of the scale factor) at $a\gg 1$. 
The condition 
$\ds\f{8\pi G}{3}\rho-\f{(mc)^2}{6}= 0$ written in the form 
$\ds\f{H^2}{\rho^0_c}\rho -\f{(mc)^2}{6}=0$ (where $H$ is the present
value of the Hubble constant) leads to the equality 
$\ds\rho_{\min}=\f{(mc)^2}{6H^2}\rho^0_c$, or in another form 
\begin{equation}
\f{\rho_{\min}}{\rho^0_c}=\f{f^2}{6}=\Omega^0_{\mbox{tot}}-1. 
\label{eq146}
\end{equation}
\textit{The field theory of gravitation appears incompatible with the
existence of the constant cosmological term leading to an unlimited
expansion of the Universe.} In fact, 
at  $a\gg 1$ it follows from equation~(\ref{eq136}):
\vspace*{3mm}
\[
\Omega^0_\Lambda < \f{f^2}{6}\ .
\]
However, this inequality is incompatible with the condition 
\[
\Omega_\Lambda ^{0}> \f{f^2}{6}\,,
\]
which is needed in order that in the present time there existed,
according to equation 
(\ref{eq137}), an accelerated expansion.
\vspace*{3mm} 

\textit{Thus the only possibility to explain, in the framework of the
RTG,  the accelerated expansion of
the Universe  observed at the present time is the existence of a
quintessence with 
$\nu >0$  or some other substance the density of which decreases with
increase of the scale factor (but not faster than  
$\mbox{const}/a^2$). The RTG excludes a possibility of the existence
both of the constant cosmological term 
$(\nu=0)$ and the  ``phantom'' expansion  $(\nu<0)$} [23].
\vspace*{3mm}

\subsection{Start and finish of the present accelerated expansion} 
\vspace*{3mm}

The most strict bounds,  
$\Omega^0_{\mbox{tot}} =1{,}018^{+0{,}013}_{-0{,}022}$\,, obtained in
the experiment 
 WMAP~[17] in the framework of $\Lambda$CDM-model with use of the
 data
 from the galaxy catalogue  SDSS and the data on supernovae SNIa
admit  within 
$1\sigma$  the value 
$\Omega^0_{\mbox{tot}}=1{,}03$.
This difference in the RTG, according to relations 
(\ref{eq138}), (\ref{eq139}),  
\textit{determines the graviton mass} 
\[
m_g=0{,}424\,m_H=1{,}6\cdot 10^{-66}\,h\ .
\]
\vspace*{3mm}
\textit{Further  we shall use for definitness namely this value of
the graviton mass.} Since by the beginning of the epoch of the
present
acceleration 
$\Omega_r\ll \Omega_m$ and  $a\gg 1$, then the start and finish of
the accelerated expansion are determined, according to 
(123), by the roots 
$x_1<1 < x_2$ of the equation $F(x)=0$, where the function  $F(x)$ is 
\[
F(x)=\f{\Omega^0_m}{x^3}-2\left (
1-\f{3\nu}{2}\right )
\f{\Omega^0_q}{x^{3\nu}}
+\f{f^2}{3}\ . 
\]
Here the value of the first root  $x_1$  is related to the red shift 
$Z_1$, corresponding to the beginning of the acceleration epoch
\vspace*{5mm}
\be
\f{1}{x_1}=\f{a_0}{a_1}
=Z_1+1\ .
\label{eq147}
\ee
The time lapsed from the beginning of the Universe expansion till the
beginning of the present acceleration can be establishied from
equation  (\ref{eq136}).
Neglecting the duration of the radiation dominated epoch and the
value of the scale factor 
а $a$ by its end, we have 
\vspace*{5mm}
\[
\tau_1\approx \f{\,1\,}{H}
\int\limits^{x_1}_{0}\!\!\f{dx}{x[\varPhi (x)]^{1/2}}=
\f{\,1\,}{H}
\int\limits^{\infty}_{Z_1+1}
\f{dy}{y\left(\Omega^0_m y^3+\Omega^0_qy^{3\nu}
-f^2/6\right )^{1/2}}\,, 
\]
\vspace*{3mm}
where
\vspace*{0.3cm}
\[
\varPhi (x) =\f{\Omega^0_m}{x^3}
+\f{\Omega^0_q}{x^{3\nu}}
-\f{f^2}{6}\,.
\]

\vspace*{0.6cm}
Here the values $\Omega^0_m=0{,}27$, 
$\Omega^0_q=0{,}73$ are assumed according to [25]. 

\vspace*{3mm}

Correspondingly, the time of the termination of the epoch of
accelerated expansion and of the passage to deceleration is 
\vspace*{0.3cm}
\[
\tau_2=\f{\,1\,}{H}
\int\limits^{x_2}_{0}
\!\!\f{dx}{x[\varPhi (x)]^{1/2}}\,,
\]
\vspace*{0.3cm}
and the present age of the Universe,   $\tau_0$, is:
\vspace*{0.3cm}
\[
\tau_0=\f{\,1\,}{H}
\int\limits^{1}_{0}
\!\!\f{dx}{x[\varPhi (x)]^{1/2}}\,. 
\]

\vspace*{0.5cm}
The physical distance passed by the light (particle horizon) by the
present moment of time is determined by the following expansion 

\vspace*{0.5cm}
\ba
&&D_{\mbox{part}}(\tau_0)
=\f{\,c\,}{H}\!\!\int\limits_1^{a_0/a_{min}}\!\!dy
\f{1}{[\Omega_r^0y^4+\Omega_m^0y^3
+\Omega_q^0y^{3\nu}-f^2/6
\times(1+y^6/2a_0^6)]^{1/2}}\simeq\nonumber \\
&&\simeq\f{2}{\s{\Omega_m^0}}\,\f{\,c\,}{H}\;.\nonumber
\ea

\vspace*{0.3cm}
This quantity determines the size of the observed Universe by the
present time. Qualitatively (without exact scales) the temporal
dependence of the scale factor, its velocity 
 $\dot a$ and  $\ddot a$ is given in the Figure. 

\begin{figure}[hbp]
$$
\includegraphics[width=78mm]{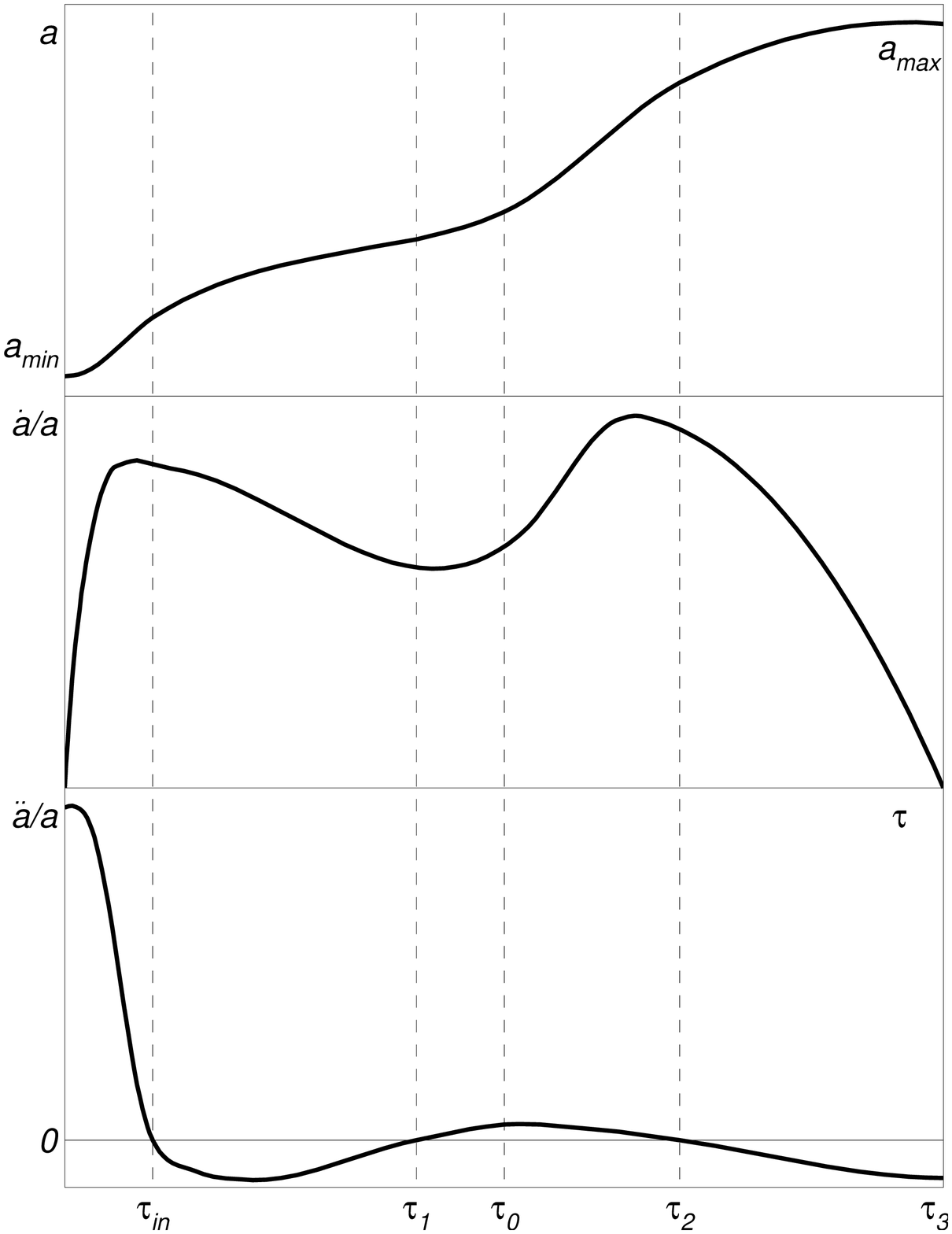}
$$
\vspace*{2mm}

{\bf{Figure}. {\sl Qualitative curves of the dependence of the scale
factor (upper part)
velocity and acceleration (lower part) dependent on time 
$\tau$. Here $\tau_{in}~=1{,}15\tau_r $. The present moment of time
is designated  $\tau_{0}$. In the beginning the scale factor grows
from its minimum value 
 $a_{min}$ with a very high acceleration which in a short time
 enough, 
 $\tau_{in}$, becomes zero.  The velocity in this period of time
 increases from the zero value up to the maximum one. The scale
 factor during this period of time changes insignificantly:
$a(\tau_{in})~= \sqrt{2} a_{min}$. Further on the expansion occurs
 with negative acceleration which becomes zero at
some moment of time 
 $\tau_{1}$. The value of the velocity drops and somewhat later than 
 $\tau_{1}$ it achieves its minimum value. 
The scale factor in this period of time continues to rise (expansion
continues). The motion with positive acceleration continues till the
moment 
$\tau_{2}$. The velocity and the scale factor increase. At 
 $\tau > \tau_{2}$ the expansion occurs again with negative
 acceleration until the time when at the momentum 
 $\tau_{3}$ the expansion stops. The  scale factor achieves its
 maximum value. On this the half-cycle is completed and everything
 repeats
 in the opposite order: the expansion epoch is changed by the
 compression epoch. For the quantity  ${\dot a}/a$
the first maximum is situated at  $a=\sqrt{3/2}~a_{min}$
($\tau \sim 0{,}76~\tau_{r}$) somewhat earlier than  $\tau_{in}$, 
in the some way as the second maximum happens prior to  $\tau_{2}$.
The minimum of  ${\dot a}/a$, contrary to this, is situated later
than 
$\tau_{1}$. This follows from the fact that the quantity 
$(d/d\tau)(\dot a/a)=(\ddot a/a)-(\dot {a}^2/a^2)$
at  ${\ddot a}= 0$ is negative.
}}
\end{figure}

\newpage
\subsection{Maximum value of the scale factor and the integral 
of~the~Universe~evolution}

The time corresponding to the end of the accelerated expansion and
the beginning of deceleration leading to the stop of expansion
strongly depends on parameter 
 $\nu$ (see the Table). 

\begin{figure}[!hbp]
\begin{center}
{\bf Table}.
The time of the beginning of the accelerated expansion of the
Universe,  $\tau_1$, that of its termination, 
$\tau_2$, and the time of the maximum expansion (oscillation
half-period) 
$\tau_{\mbox{{\small max}}}$ [bill. years].
\\
\begin{tabular}{|p{25mm}|p{25mm}|p{25mm}|p{25mm}|}
\hline
\multicolumn{1}{|c|}{$\nu$}
& \multicolumn{1}{|c|}{$\tau_1$} &
\multicolumn{1}{|c|}{$\tau_2$} &
\multicolumn{1}{|c|}{$\tau_{\mbox{{\small max}}}$}     \\
\hline\hline
$\quad\nu = 0{,}05$ &\quad  7{,}0 - 8{,}2 &\quad  980 - 1080 &\quad
1220 - 1360 \\ \hline $\quad\nu = 0{,}10$ &\quad  7{,}0 - 8{,}2
&\quad  440 - 485  &\quad  620 - 685   \\ \hline 
$\quad\nu = 0{,}15$ &\quad  7{,}1 - 8{,}3 &\quad  275 - 295  &\quad
430 - 460   \\ \hline 
$\quad\nu = 0{,}20$ &\quad  7{,}1 - 8{,}3 &\quad  190 - 205  &\quad
325 - 347   \\ \hline 
$\quad\nu = 0{,}25$ &\quad  7{,}2 - 8{,}5 &\quad  142 - 149  &\quad
263 - 280   \\ \hline 
$\quad\nu = 0{,}30$ &\quad  7{,}5 - 8{,}7 &\quad  109 - 113  &\quad
227 - 235   \\ \hline 
\end{tabular}
\end{center}
\end{figure}

The scale-factor, corresponding to the stop of the expansion, 
 $x_{\max}$, is determined by the root of equation 
 (\ref{eq136}) and at small  $\nu$ is, with a good accuracy, 
 \be
x_{\max}\simeq \Bigl(\f{6\Omega^0_q}{f^2}\Bigr)^{1/3\nu}
=\Bigl(\f{\Omega_q^0}{\Omega_{\tot}^0-1}\Bigr)^{\!1/3\nu}\!.
\label{eq148}
\ee 
Substituting the value of  $a_0$ from equation 
(\ref{eq145}) into this expression, we find 
\[
a_{\max}^4=\f{1}{\Omega_r^0}\Bigl(\f{f^2}{6}\Bigr)^{2/3}\Bigl(\f{2\pi
}{3} \f{G\rho_{max}}{H^2}\Bigr)^{1/3}
\Bigl(\f{\Omega_q^0}{\Omega_{\tot}^0-1}\Bigr)^{\!4/3\nu}\!.
\]
Taking into account this equality and that the integral of motion is 
\[
E=\f{(mc)^2}{8a_{\max}^4}\,,
\]
we obtain 
\[
E=\f{(mc)^2}{8}\Omega_r^0\Bigl(\f{6}{f^2}\Bigr)^{2/3}\Bigl(\f{3}{2\pi
}\f{H^2}
{G\rho_{max}}\Bigr)^{1/3}
\Bigl(\f{\Omega_{\tot}^0-1}{\Omega_q^0}\Bigr)^{\!4/3\nu}\!.
\]
It is evident from here that the integral of motion of the Universe
evolution is a very small quantity. Making use of the expression for 
$x_{\max}$,  it is easy to determine the relative acceleration of the
attraction in the moment of the stop of expansion:
\[
\frac{\,\ddot a\,}{a} \sim -\frac{\,\nu\,}{4} \left(
\frac{m_{g}c^{2}}{\hbar}
\right)^{2}\,,
\]
and therefore the scalar curvature $R$ is 
\[
R=\f{3\nu}{2c^2}\Bigl(\f{m_g c^2}{\hbar}\Bigr)^2\,.
\]
It is essential that the relative minimum value of the density 
$(\rho_{\min}/\rho^0_c)$,  corresponding to the maximum of expansion
depends on the value of 
 $(\Omega^0_{tot}-1)$ only , i.e.  on the graviton mass 
 (see (\ref{eq139}), (\ref{eq140})). At $(\Omega^0_{tot}=1{,}02)$ 
 the value of $\rho_{\min}$ is quite large and even exceeds very much
 the 
 \textit{present density of radiation}. In paper [24] the authors
 proceeded from the present age of the Universe, 
$(13{,}7\pm 0{,}2)\cdot
10^9$ years given in [25, 26]. This quantity is calculated in 
[25, 26] basically in the 
$\Lambda\!$\textit{TDі-model}. 
It is very important that the recent observation of  SN1a
[27, 28] in the range $Z\gtrsim 1$ can give  \textit{a
straightforward information} about the beginning of the present
acceleration. Such data were obtained in the excellent paper by
A.~Ries et al [28] according to which the deceleration was changed by
the present acceleration at the following values of the red shift 
\[
Z=0{,}46\pm 0{,}13\ .
\]
This result conforms with the presented picture of evolution. It
enables to directly determine the value 
 $x_1$ (see (\ref{eq147})) and to precise the admissible range of the
 cosmological parameters \footnote{Let us notice that the distance to
 supernovae  $(D_L)$,
determined from the relation  $F=L/4\pi D_L^2$ 
(where $L$ and $F$ are the luminosity of the standard SNla and
observed flow, respectively), is expressed via cosmological
parameters of the RTG by the relation 
\[
D_L=\f{c}{H}(Z+1)\int\limits_1^{1+Z}
\!\!\Bigl[\Omega_m^0 y^3+\Omega_q^0 y^{3\nu}-
\f{f^2}{6}\Bigr]^{-1/2}dy.
\]
}. 

The expansion up to the maximum value of the scale parameter and its
consequent compression lead to the oscillatory character of the
Universe evolution. The idea of the oscillatory character of the
Universe evolution was repeatedly advanced earlier proceeding mainly
from the philosophical considerations 
(see, e.g., [29]~-- [31]). Such a regime could, in principle, be
expected in the closed Friedmann model with 
 $\Omega_{\mbox{tot}}>1$. However, firstly,  the insurmountable
 difficulty
 related to the passage through the cosmological singularity and,
 secondly, coniderations related to the growth of entropy from cycle
 to cycle  [31] do not allow this. 

It is necessary to emphasize that in the framework of the
Hilbert--Einstein equations the flat Universe cannot be
oscillatory \footnote{Paper [32] on the cyclic evolution of the
Universe is erroneous because the ``solution'' given in it is not in
fact a solution to the basic system of the Hilbert--Einstein
equations, which can be checked by the direct substitution. Paper
[33] is also erroneous because the system of equations (3), (17) and
(18) of this article is internally contradictory.}.
 These difficulties for the infinite Universe are eliminated in the
 RTG. Since singularities are absent from the RTG the Universe could
 exist an infinite time during which the interaction occured among
 its domains and this led to homogeneity and isotropy of the Universe
 with some structure of inhomogeneity which we did not take into
 account for the sare of simplicity. 

In the above-mentioned approximation  $x_{max}$ is related to the
scale factor $(x_2)$ corresponding to the termination of the
accelerated expansion by the relation
\[
x_2=\left
(1-\f{\,3\,}{2}\nu\right )^{\!\!{1/3\nu}}\!\!\!\!\!\!\!\!\cdot
x_{\max}\approx \f{1}{\s e}x_{\max}\,.
\]

The time corresponding to the stop of the expansion (half-period of
oscillation) with chosen in [24] value of the graviton mass 
 $m_g=0{,}49\,m_H$ 
is about  $1300\cdot 10^9$years for 
$\nu=0{,}05$, near $650\cdot 10^9$years for  $\nu=0,10$ 
and $270\cdot
10^9$years for $\nu=0{,}25$. 

The attractivness of the oscillatory evolution of the Universe is
mentioned in the recent paper~[34]. The oscillatory regime is
realized by the price of introducing a scalar field interacting with
the
substance and use of the extra dimensions. Some important
consideration were advanced that the phase of the accelerated
expansion promotes the entropy conservation in the repeating cycles
of the evolution. In the RTG the oscillatory character of the
Universe evolution is achieved as a result of introducing of the only
massive gravitational field as a physical field generated by the
total energy-momentum tensor in the Minkowski space.\\ 

In conclusion the authors express their gratitude to V.V.Kiselev, Yu.M.~Loskutov, 
V.A~Petrov and
N.E~Turyin for valuable discussions.\\ 

\renewcommand{\refname}

\end{document}